\documentclass[aps,prl,preprintnumbers,showpacs,nofootinbib]{revtex4}
\usepackage[utf8]{inputenc}
\usepackage[english]{babel}
\usepackage{amsmath}
\usepackage{amsfonts}
\usepackage{amssymb}
\usepackage{epsfig}
\usepackage{graphics,psfrag,rotating}
\usepackage{graphicx}
\usepackage{dcolumn}
\usepackage{bm}
\usepackage{epstopdf}
\usepackage{color}
\usepackage[usenames,dvipsnames,svgnames]{xcolor}
\usepackage[colorlinks=true,
            linkcolor=red,
            urlcolor=gray,
            citecolor=blue]{hyperref}
  \usepackage{hyperref}

\newcommand{\lsim}   {\mathrel{\mathop{\kern 0pt \rlap
{\raise.2ex\hbox{$<$}}}
 \lower.9ex\hbox{\kern-.190em $\sim$}}}
\newcommand{\gsim}   {\mathrel{\mathop{\kern 0pt \rlap
{\raise.2ex\hbox{$>$}}}
\lower.9ex\hbox{\kern-.190em $\sim$}}}

\def\3nab{\tilde{\nabla}}

\def\hsp5{\hspace{5mm}}

\def\case#1/#2{\textstyle\frac{#1}{#2}}

\def\ber {\begin{eqnarray}}
\def\eer {\end{eqnarray}}
\def\bea {\begin{eqnarray}}
\def\eea {\end{eqnarray}}

\def\bc {\begin{center}}
\def\ec {\end{center}}
\def\case#1/#2{\frac{#1}{#2}}

\newcommand{\bw}{\begin{widetext}}
\newcommand{\ew}{\end{widetext}}

\newcommand{\be}{\begin{equation}}
\newcommand{\bse}{\begin{subequation}}
\newcommand{\ese}{\end{subequation}}
\newcommand{\ee}{\end{equation}}
\newcommand{\eei}{\end{eqnarray}\indent\indent}
\newcommand{\ba}{\begin{array}}
\newcommand{\ea}{\end{array}}
\newcommand{\bal}{\begin{eqnarray}}
\newcommand{\eal}{\end{eqnarray}}

\def\case#1/#2{\textstyle\frac{#1}{#2} }

\newcommand{\bver}{\begin{verbatim}}
\newcommand{\ever}{\end{verbatim}}

\begin{document}

%\preprint{APS/123-QED}

\title{A Bayesian PINN Framework for Barrow–Tsallis Holographic Dark Energy with Neutrinos: Toward a Resolution of the Hubble Tension}
\author{
 Muhammad Yarahmadi$^{1}$\footnote{Email: yarahmadimohammad10@gmail.com}
   , Amin Salehi$^{1}$ 
}
\affiliation{Department of Physics, Lorestan University, Khoramabad, Iran}

\date{\today}% It is always \today, today,

\begin{abstract}
We investigate the Barrow–Tsallis Holographic Dark Energy (BTHDE) model using both traditional Markov Chain Monte Carlo (MCMC) methods and a Bayesian Physics-Informed Neural Network (PINN) framework, employing a range of cosmological observations. Our analysis incorporates data from Cosmic Microwave Background (CMB), Baryon Acoustic Oscillations (BAO), CMB lensing, Cosmic Chronometers (CC), and the Pantheon+ Type Ia supernova compilation. We focus on constraining the Hubble constant \( H_0 \), the nonextensive entropy index \( q \), the Barrow exponent \( \Delta \), and the Granda–Oliveros parameters \( \alpha \) and \( \beta \), along with the total neutrino mass \( \Sigma m_\nu \). The Bayesian PINN approach yields more precise constraints than MCMC, particularly for \( \beta \), and tighter upper bounds on \( \Sigma m_\nu \). The inferred values of \( H_0 \) from both methods lie between those from Planck 2018 and SH$_0$ES (R22), alleviating the Hubble tension to within \( 1.3\sigma \)–\( 2.1\sigma \) depending on the dataset combination. Notably, the Bayesian PINN achieves consistent results across CC and Pantheon+ datasets, while maintaining physical consistency via embedded differential constraints. The combination of CMB and late-time probes leads to the most stringent constraints, with \( \Sigma m_\nu < 0.114 \) eV and \( H_0 = 70.6 \pm 1.35 \) km/s/Mpc. These findings suggest that the BTHDE model provides a viable framework for addressing cosmological tensions and probing modified entropy scenarios, while highlighting the complementary strengths of machine learning and traditional Bayesian inference in cosmological modeling.
\end{abstract}


\maketitle

\section{Introduction}
\label{intro}
The cosmological constant \( \Lambda \) remains the simplest and most widely studied candidate for dark energy (DE), successfully accounting for the observed acceleration of the universe within the standard cosmological framework, known as the concordance \( \Lambda \)CDM model~\cite{yur,prob,jass}. Despite its success, mounting observational evidence has revealed significant tensions within this model, particularly between early- and late-universe measurements of the Hubble constant and other cosmological parameters~\cite{yur,prob,cm,tt,carol2}. In response, a variety of alternative approaches have been proposed, broadly classified into two categories: modifications of general relativity, and dynamical DE models with evolving energy densities and equation of state (EoS) parameters~\cite{two}. Among these, several scalar field-based models have garnered considerable attention, including quintessence, K-essence, phantom, quintom, and tachyon models~\cite{rat,caa,fen,sahh,caai}.

One particularly intriguing direction is the holographic dark energy (HDE) paradigm, which is rooted in the holographic principle, a concept inspired by quantum gravity that relates the degrees of freedom of a volume of space to those on its boundary~\cite{sus2,sus3,sus4,moo,moo1,moo2,moo3,moo4}. Within this framework, the entropy associated with the horizon plays a central role, and the Bekenstein--Hawking entropy \( S = A/4G \), where \( A \) is the area of the apparent horizon, was among the first formulations connecting thermodynamics with gravitational systems~\cite{bh2}. 

Building on this foundation, various generalized entropy forms have been introduced to describe different physical systems, especially those exhibiting nonextensive or quantum gravitational behavior. These include, but are not limited to, Tsallis, Kaniadakis, and Barrow entropies~\cite{tes,reni,say,maj,k1,k2}. Such generalized entropy frameworks offer promising avenues for modifying the standard HDE model, enabling a deeper understanding of late-time cosmic acceleration and its connection to fundamental physics.

One of the most pressing challenges in contemporary cosmology is the so-called Hubble tension —a statistically significant discrepancy between measurements of the Hubble constant \( H_0 \), the current rate of cosmic expansion. This tension arises from two primary and independently calibrated measurement techniques.

On one side, early-universe observations, primarily those from the \textbf{Planck} satellite analyzing the cosmic microwave background (CMB) under the assumption of the $\Lambda$CDM model, yield a lower value of the Hubble constant:

$ H_0 = 67.4 \pm 0.5 \, \text{km}\,\text{s}^{-1}\,\text{Mpc}^{-1} $ \quad \text{\cite{Planck18}}. In contrast, late-universe observations based on local distance ladders, such as Cepheid-calibrated Type Ia supernovae from the SH0ES (Supernovae, H$_0$, for the Equation of State) project using the Hubble Space Telescope, report a significantly higher value: $ H_0 = 73.04 \pm 1.04 \, \text{km}\,\text{s}^{-1}\,\text{Mpc}^{-1} $ \quad \text{\cite{riess2022}}. The disagreement between these measurements currently exceeds the \( 4.4\sigma \) level, raising concerns about possible unknown systematics or the need for new physics beyond the standard $\Lambda$CDM paradigm. Several independent observational projects have contributed further \( H_0 \) estimates, often yielding intermediate values but reinforcing the tension:

\begin{itemize}
	\item \textbf{CCHP (TRGB)}: \( H_0 = 69.6 \pm 0.8 \pm 1.7 \, \text{km}\,\text{s}^{-1}\,\text{Mpc}^{-1} \) \cite{FreedmanTRGB20}
	\item \textbf{HST (Miras)}: \( H_0 = 72.7 \pm 4.6 \, \text{km}\,\text{s}^{-1}\,\text{Mpc}^{-1} \) \cite{HuangMiras19}
	\item \textbf{H0LiCOW (strong lensing)}: \( H_0 = 73.3^{+1.7}_{-1.8} \, \text{km}\,\text{s}^{-1}\,\text{Mpc}^{-1} \) \cite{Wonglens19}
	\item \textbf{H0LiCOW (updated)}: \( H_0 = 75.3^{+3.0}_{-2.9} \, \text{km}\,\text{s}^{-1}\,\text{Mpc}^{-1} \) \cite{Weilens20}
	\item \textbf{Baxter (CMB lensing)}: \( H_0 = 73.5 \pm 5.3 \, \text{km}\,\text{s}^{-1}\,\text{Mpc}^{-1} \) \cite{BaxterCMBlens20}
\end{itemize}

\vspace{1em}

To address this tension, several extensions of the standard model have been proposed, particularly those that modify the dark energy sector. One promising avenue is the \emph{Barrow–Tsallis Holographic Dark Energy} (BTHDE) model, which generalizes holographic dark energy using ideas from non-extensive entropy frameworks: Barrow entropy and Tsallis statistics.
 By allowing dynamical deviations from conventional dark energy behavior, the BTHDE model naturally adjusts the expansion rate of the universe in the late-time epoch without conflicting with early-universe constraints. This flexibility allows BTHDE to interpolate between Planck and SH0ES measurements, effectively reducing the statistical significance of the Hubble tension.
Furthermore, by introducing an additional contribution to the energy budget that evolves differently than a cosmological constant, the BTHDE model modifies the inferred distance-redshift relation, which directly affects the calibration of standard candles like supernovae. As a result, the model offers a viable and physically motivated resolution to the Hubble discrepancy within a Bayesian framework.

Barrow~\cite{5} recently investigated the impact of quantum gravitational effects on black hole (BH) horizons. He proposed that such effects could induce complex, fractal-like structures on the event horizon, leading to a modification of the standard entropy-area relation. The resulting expression for the black hole entropy, often referred to as Barrow entropy, is given by	$S_B = \left( \frac{A}{A_0} \right)^{1 + \frac{\Delta}{2}}$ where \( A \) denotes the horizon area, \( A_0 \) is the Planck area, and \( \Delta \) is the deformation parameter quantifying the degree of quantum gravitational corrections. In the limit \( \Delta = 0 \), one recovers the standard Bekenstein--Hawking entropy, while \( \Delta = 1 \) corresponds to maximal deformation. Throughout this work, we adopt natural units, setting \( \hbar = c = k_B = 1 \). The cosmological implications of Barrow entropy have been the subject of extensive investigation. A holographic dark energy (HDE) model based on $S_B$ was proposed in Ref.~\cite{6}, successfully describing the transition between matter- and dark-energy-dominated eras. Observational constraints on this Barrow HDE model were examined in Ref.~\cite{7}, where consistency with current cosmological data was demonstrated. Moreover, a modified gravitational framework incorporating Barrow entropy was formulated in Ref.~\cite{8}, where the extended field equations give rise to an effective dark energy component. The validity of the generalized second law of thermodynamics in a Barrow entropy setting was studied in Ref.~\cite{9}, and further developments, both theoretical and observational, can be found in Refs.~\cite{10,11}.

Although originally introduced as a theoretical construct, the Barrow entropy formalism is now supported by a growing body of literature, highlighting its potential relevance to quantum gravity and early-universe physics. Its corrections to the entropy-area law naturally propagate into the Friedmann equations, modifying the dynamics of cosmic expansion and motivating a thorough re-examination of standard cosmological scenarios.

In this context, the combined Barrow and Tsallis entropy formalisms offer a compelling theoretical framework to address outstanding issues in cosmology. Barrow entropy, which emerges from quantum gravitational deformations of spacetime geometry, introduces the deformation parameter \( \Delta \), thereby altering the gravitational dynamics. Tsallis entropy, on the other hand, generalizes the Boltzmann--Gibbs formalism to account for systems with long-range correlations and nonextensivity, characterized by the entropic index \( q \). When these two frameworks are jointly employed and extended to include logarithmic corrections, they lead to modified Friedmann equations with effective dark energy terms that can influence the intermediate redshift dynamics.

Such models not only provide a potential resolution to the longstanding \( H_0 \) tension between early- and late-universe measurements of the Hubble parameter, but also offer deeper insight into the nature of dark energy, cosmic inflation, and the thermodynamics of the apparent horizon. The presence of tunable parameters such as \( \Delta \) and \( q \) enhances the flexibility of these models, allowing them to be tested against current and forthcoming observational data, and positioning them as valuable tools in the quest for a unified and quantum-informed cosmological theory.
The Barrow--Tsallis Holographic Dark Energy (BTHDE) model emerges from a compelling synthesis of Barrow's entropy deformation, motivated by quantum gravitational effects, and Tsallis' nonextensive statistical mechanics. In this framework, the standard holographic dark energy (HDE) density is modified by incorporating both the deformation parameter \( \Delta \), which accounts for fractal features of the cosmic horizon due to quantum fluctuations~\cite{5}, and the Tsallis entropic index \( q \), which captures the nonextensive behavior of gravitational systems with long-range interactions~\cite{Tsallis}. The resulting energy density deviates from the standard HDE form, yielding a more flexible model capable of describing a broader range of cosmic expansion histories. When implemented with an appropriate infrared (IR) cutoff, such as the Granda--Oliveros scale, the BTHDE model introduces new dynamical degrees of freedom that influence the late-time acceleration and can reconcile tensions between early- and late-universe observations. Recent studies have shown that this model not only accommodates the observed accelerated expansion, but also provides a promising approach to addressing the Hubble tension and other anomalies in cosmological data. Owing to its solid theoretical foundation and observational viability, the BTHDE framework serves as a unified and physically motivated modification of dark energy models, bridging the gap between quantum gravity, thermodynamics, and late-time cosmology.
In recent years, Physics-Informed Neural Networks (PINNs) have emerged as a powerful framework for solving differential equations and modeling physical systems by embedding physical laws directly into the training process of neural networks~\cite{Raissi2019, Lu2020, Zhang2020, Wang2022}. By incorporating the governing equations—such as the Friedmann equations in cosmology—into the loss function, PINNs ensure that the learned solutions are consistent with the underlying physics. This makes them especially suitable for modeling cosmological dynamics where analytic solutions are not available or where observational data are sparse and noisy. To enhance the robustness and interpretability of PINNs, the Bayesian PINN framework extends the deterministic PINN formulation by introducing uncertainty quantification through probabilistic inference~\cite{Yang2021b}. In this approach, Bayesian techniques are employed to infer posterior distributions over model parameters and solutions, allowing for a principled treatment of observational errors and model uncertainties. When applied to cosmology, Bayesian PINNs offer a novel data-driven tool for reconstructing cosmic histories, estimating cosmological parameters, and propagating uncertainties from observations to theoretical predictions. This approach is particularly advantageous in the context of complex dark energy models like BTHDE, where traditional numerical techniques may struggle to balance accuracy, interpretability, and physical consistency. 

\subsection*{Motivation for Using Bayesian PINNs in the BTHDE Model with Neutrinos}

The incorporation of massive neutrinos and a Bayesian Physics-Informed Neural Network (PINN) framework into the Barrow–Tsallis Holographic Dark Energy (BTHDE) model is motivated by both physical and methodological considerations. 

\vspace{0.5em}
\noindent
\textbf{1. Physical Motivation: Neutrinos in Late-Time Cosmology}

Neutrinos play a subtle yet important role in cosmological evolution \cite{Y5, Y6, Sadjadi, Sadjadi0, Sadjadi1, Anari, Wetterich, Santiago, Salehi}. While relativistic in the early universe, they become semi-relativistic at late times and contribute to the total energy density via the density parameter \( \Omega_\nu \). Their impact is especially relevant for models aiming to reconcile early- and late-universe measurements of the Hubble parameter \( H_0 \), as their free-streaming nature can affect the expansion rate and structure formation.

In the context of the BTHDE model, the presence of neutrinos adds an additional physical degree of freedom that can shift the background expansion history and influence the integrated luminosity distance and Hubble rate \( H(z) \). Since BTHDE introduces a modified dark energy density governed by entropy deformation parameters \( q \) and \( \Delta \), the interplay with neutrinos may further enable the model to interpolate between conflicting cosmological datasets such as Planck (CMB) and SH0ES (SNe Ia).

\vspace{0.5em}
\noindent
\textbf{2. Methodological Motivation: Why Use a Bayesian PINN?}

Traditional cosmological analyses often rely on parametric models and numerical integration of the Friedmann equations. However, the BTHDE model involves a nonlinear differential equation with parameters that cannot be trivially fit due to degeneracies and non-analytic dependencies, especially when neutrino contributions are included.

To address this, a Bayesian PINN provides a powerful alternative by integrating physical laws directly into the training of a neural network \cite{Yang2021b}. This approach offers several distinct advantages:

\begin{itemize}
	\item \textbf{Physics-Constrained Learning:} The PINN learns a continuous solution \( H(z) \) that satisfies the BTHDE-modified Friedmann equation across redshift, ensuring consistency with underlying physical laws rather than discrete interpolation.
	
	\item \textbf{Bayesian Inference:} Through Markov Chain Monte Carlo (MCMC) sampling embedded in the training process, the Bayesian PINN yields full posterior distributions for parameters like \( \Omega_\nu, q, \Delta, \alpha, \beta \), and \( H_0 \), providing not just point estimates but credible intervals.
	
	\item \textbf{Uncertainty Quantification:} Using techniques like Monte Carlo dropout or Bayesian layers, the PINN naturally estimates epistemic uncertainties in \( H(z) \) predictions, which is critical in high-precision cosmology ~\cite{gal2016dropout}.
	
	\item \textbf{Handling Observational Data:} The Bayesian PINN incorporates observational constraints (e.g., Pantheon+ SNe Ia, BAO, or cosmic chronometers) into its loss function via a data likelihood (e.g., $\chi^2$), enabling simultaneous model training and parameter estimation.
\end{itemize}

\vspace{0.5em}
\noindent
\textbf{3. Resolving Cosmological Tensions with BTHDE + Neutrinos + Bayesian PINN}

By combining the entropy-deformed dark energy component of BTHDE with massive neutrino contributions, and analyzing the system via a Bayesian PINN, one obtains a flexible, data-driven model that adheres to fundamental physics. This integrated framework has the potential to:

\begin{itemize}
	\item Reduce the Hubble tension by adjusting the late-time expansion history;
	\item Improve the estimation of the sum of neutrino masses \( \sum m_\nu \), which is indirectly encoded through \( \Omega_\nu \) via the relation \( \sum m_\nu = 94 \, h^2 \Omega_\nu \, \text{eV} \);
	\item Quantify degeneracies and correlations among cosmological parameters with principled uncertainty estimates;
	\item Provide a robust surrogate model for complex differential equations in cosmology.
\end{itemize}

Overall, the Bayesian PINN approach offers a principled and powerful tool to extract cosmological insights from complex models like BTHDE, especially in the presence of massive neutrinos and under observational tension in parameters like \( H_0 \).
%%%%%%%%%%%%%%%%%%%%%%%%%%%%%%%%%%%%%%%%%%%%%%
\section{Model Framework}
Traditional holographic dark energy (HDE) models are rooted in the Bekenstein--Hawking entropy, which presupposes a smooth and differentiable spacetime manifold. This leads to the well-known entropy-area relation \( S = A / (4G) \), where \( A \) is the area of the cosmic horizon and \( G \) denotes Newton's gravitational constant. While this expression is reliable in the semi-classical regime, it becomes inadequate near the Planck scale, where quantum gravitational effects and non-classical statistical behavior may emerge.

Two major generalizations of the standard entropy-area relation have recently gained prominence:

\begin{itemize}
	\item \textbf{Barrow entropy}, which incorporates quantum gravitational deformations, characterized by a dimensionless parameter \( \Delta \in [0,1] \), encoding fractal-like deviations from smooth horizon geometry;
	\item \textbf{Tsallis entropy}, which generalizes Boltzmann--Gibbs statistics via a non-additive entropy measure governed by a real parameter \( q \), accounting for non-extensive behavior of gravitational degrees of freedom.
\end{itemize}

The interplay between these two corrections yields a unified framework—the \textbf{Barrow--Tsallis Holographic Dark Energy (BT-HDE)} model—which captures both quantum-gravitational and non-extensive thermodynamic features. This model provides a flexible generalization of HDE and may offer new insights into the late-time acceleration of the Universe.

\section{Generalized Barrow--Tsallis Entropy}

In this section, we construct a unified entropy measure that incorporates both the Barrow and Tsallis formalisms.
\subsection{Barrow Entropy}
\label{subsec:barrow_entropy}

Barrow proposed that the underlying quantum-gravitational structure of spacetime may induce a fractal-like deformation of the horizon geometry, thereby modifying the classical Bekenstein--Hawking entropy-area law \cite{Barrow2020}. Such quantum fluctuations near the Planck scale could effectively alter the holographic equipartition of degrees of freedom. This leads to a generalized entropy expression given by:
\begin{equation}
	S_{\mathrm{B}} = \eta \left( \frac{A}{A_0} \right)^{1 + \frac{\Delta}{2}},
	\label{eq:barrow_entropy}
\end{equation}
where \( A = 4\pi L^2 \) denotes the area of the horizon with infrared (IR) cutoff scale \( L \), \( A_0 \equiv \ell_{\mathrm{P}}^2 \) is the Planck area, and \( \eta \) is a dimensionless proportionality constant. The parameter \( \Delta \in [0,1] \) encapsulates the degree of quantum-gravitational deformation. When \( \Delta = 0 \), one recovers the standard Bekenstein--Hawking entropy, thus ensuring consistency with semiclassical gravity in the appropriate limit.

\subsection{Tsallis Entropy}
\label{subsec:tsallis_entropy}

Tsallis introduced a non-additive generalization of the Boltzmann--Gibbs entropy tailored for systems characterized by long-range interactions, memory effects, or multifractal boundary conditions \cite{Tsallis, Sheykhi2012, Sheykhi2019, Moradpour2018, Saridakis2018}. Within the framework of gravitational systems—where traditional extensivity may be violated—this approach yields a generalized entropy expression of the form:
\begin{equation}
	S_{\mathrm{T}} = \alpha A^{\frac{1 - q}{2} + 1},
	\label{eq:tsallis_entropy}
\end{equation}
in which \( q \in \mathbb{R} \) denotes the Tsallis non-extensive parameter and \( \alpha \) is a normalization constant with appropriate units. The limit \( q \rightarrow 1 \) smoothly reproduces the extensive Boltzmann--Gibbs entropy, thereby ensuring thermodynamic consistency in the classical regime.

\subsection{Barrow--Tsallis Unified Entropy}
\label{subsec:bt_entropy}

To simultaneously incorporate quantum-gravitational corrections (captured by Barrow's exponent \( \Delta \)) and non-extensive thermodynamic effects (encoded in Tsallis's \( q \)-parameter), we propose a unified entropy framework. This construction proceeds by modifying the area-dependence of the Tsallis entropy in accordance with Barrow's deformation, effectively substituting:
\begin{equation}
A \rightarrow A^{1 + \frac{\Delta}{2}}.
\end{equation}
Accordingly, the generalized entropy becomes:
\begin{equation}
	S = \gamma \left( A^{1 + \frac{\Delta}{2}} \right)^{\frac{3 - q}{2}} = \gamma A^{\left(1 + \frac{\Delta}{2} \right)\frac{3 - q}{2}},
	\label{eq:bt_entropy}
\end{equation}
where \( \gamma \) is a normalization constant that can absorb the dimensional scaling. To express the entropy in a manifestly dimensionless form relative to the Planck scale, we define:
\begin{equation}
	S = \gamma \left( \frac{A}{A_0} \right)^{\left(1 + \frac{\Delta}{2} \right)\frac{3 - q}{2}}.
	\label{eq:bt_entropy_dimless}
\end{equation}

Equation~\eqref{eq:bt_entropy_dimless} defines what we term the \textbf{generalized Barrow--Tsallis entropy}. This expression encapsulates both the fractal deformation of the horizon geometry due to quantum-gravitational phenomena and the non-extensive statistical nature of gravitational systems. As shown in Refs.~\cite{Barrow2020, Tsallis, javed2024, gennaro2022}, such entropy generalizations serve as a compelling foundation for modified holographic dark energy models. In particular, Eq.~\eqref{eq:bt_entropy_dimless} underlies the construction of the Barrow--Tsallis Holographic Dark Energy (BT-HDE) model, from which the corresponding energy density can be derived. This entropy-based framework thus provides a natural platform to investigate cosmological phenomena in the presence of both quantum spacetime corrections and generalized thermodynamic effects.

\subsection{Consistency Checks}

It is crucial to verify that the generalized Barrow--Tsallis entropy reduces to the well-known entropy forms under appropriate limits. Such consistency checks validate the robustness and physical plausibility of the proposed model.

\paragraph{(i) Bekenstein--Hawking limit (\(q = 1\), \(\Delta = 0\)):}

Substituting these values into the generalized entropy expression \eqref{eq:bt_entropy_dimless}, we obtain:
\begin{equation}
	S = \gamma \left( \frac{A}{A_0} \right)^{\left(1 + \frac{0}{2} \right)\frac{3 - 1}{2}} 
	= \gamma \left( \frac{A}{A_0} \right)^1,
\end{equation}
which reproduces the standard Bekenstein--Hawking entropy:
\begin{equation}
	S \propto A.
\end{equation}
This ensures compatibility with the semi-classical thermodynamics of smooth black hole horizons.

\paragraph{(ii) Barrow limit (\(q = 1\), \(\Delta \neq 0\)):}

For this case, the non-extensive effects vanish, but quantum gravitational deformations remain. The entropy becomes:
\begin{equation}
	S = \gamma \left( \frac{A}{A_0} \right)^{\left(1 + \frac{\Delta}{2} \right)\frac{3 - 1}{2}} 
	= \gamma \left( \frac{A}{A_0} \right)^{1 + \frac{\Delta}{2}},
\end{equation}
which is precisely the Barrow entropy. This consistency highlights that the model retains the effect of horizon deformation in the absence of non-extensive statistics.

\paragraph{(iii) Tsallis limit (\(\Delta = 0\), \(q \neq 1\)):}

In this limit, quantum gravitational corrections vanish, but the entropy still incorporates statistical non-extensivity:
\begin{equation}
	S = \gamma \left( \frac{A}{A_0} \right)^{\left(1 + \frac{0}{2} \right)\frac{3 - q}{2}} 
	= \gamma \left( \frac{A}{A_0} \right)^{\frac{3 - q}{2}}.
\end{equation}
This recovers the entropy profile derived from Tsallis non-extensive thermodynamics, applicable to systems with long-range interactions or memory effects, such as gravitational systems.

These limiting cases confirm that the generalized entropy expression \eqref{eq:bt_entropy_dimless} represents a consistent and encompassing framework that recovers known entropy formulas in appropriate limits.

\subsection{Physical Interpretation of Parameters}

The Barrow--Tsallis generalized entropy introduces a parameter space with distinct physical implications for spacetime thermodynamics and cosmology.

\begin{itemize}
	\item \textbf{Quantum-gravitational deformation parameter \(\boldsymbol{\Delta}\):} This dimensionless parameter encapsulates the degree of deviation from the classical horizon smoothness, arising from potential quantum fluctuations at Planckian scales. Geometrically, it quantifies the fractal-like deformation of the spacetime horizon and modifies the scaling of entropy with area. The limiting values
	\(\Delta = 0\) and \(\Delta = 1\) correspond to a smooth classical geometry and a maximally deformed (fully fractal) horizon, respectively. Physically, nonzero \(\Delta\) introduces higher-order quantum corrections to the entropy that could emerge in theories of quantum gravity such as Loop Quantum Gravity or Causal Dynamical Triangulations.
	
	\item \textbf{Non-extensivity index \(\boldsymbol{q}\):} The parameter \(q\) measures the degree of non-additivity in the underlying statistical mechanics of gravitational degrees of freedom. Systems governed by gravity naturally exhibit non-local correlations, making the Boltzmann--Gibbs framework insufficient. The Tsallis index captures deviations from extensivity: \(q = 1\) recovers standard thermodynamics, while \(q \neq 1\) signifies the presence of long-range interactions, non-Markovian dynamics, or multifractal phase-space structures. A value \(q > 1\) corresponds to sub-extensive behavior, whereas \(q < 1\) reflects super-extensivity.
	
	\item \textbf{Normalization coefficient \(\boldsymbol{\gamma}\):} The prefactor \(\gamma\) is a dimensionful constant that encapsulates the microscopic physics underpinning the entropy. It generally depends on the choice of fundamental units and may involve the Planck scale, dimensional constants from the gravitational action, or couplings inherited from quantum gravity candidates. Although \(\gamma\) does not directly affect the functional form of entropy scaling, it plays a crucial role in matching the model with observational data through the holographic energy density.
	
\end{itemize}

In the context of holographic cosmology, the combination of \(\Delta\) and \(q\) introduces a rich phenomenological landscape, where departures from classical behavior are tightly controlled by their respective values. This allows for a more nuanced understanding of dark energy dynamics and the late-time acceleration of the Universe, particularly when tested against cosmic chronometers, Type Ia supernovae, and baryon acoustic oscillation data.

The thermodynamic foundation provided by the Barrow--Tsallis entropy offers a natural route to constructing modified holographic dark energy models, bridging quantum gravity insights with cosmological observables.

\section{Energy Density of the BT-HDE Model}

According to the holographic principle, the dark energy density is determined by the entropy and the infrared (IR) cutoff. In the standard HDE model:
\begin{equation}
	\rho_D = 3c^2 M_p^2 L^{-2},
\end{equation}
where \(c^2\) is a dimensionless constant, \(M_p\) is the reduced Planck mass, and \(L\) is the IR cutoff scale (often taken as the future event horizon or the Hubble horizon).

In the BT-HDE framework, due to the modified entropy-area scaling, the energy density is also modified and depends explicitly on \(q\) and \(\Delta\), and the corresponding expression will be derived using the generalized entropy \eqref{eq:bt_entropy_dimless}.

\subsection{Derivation of the Modified Energy Density}

We begin from the first law of thermodynamics applied to the apparent horizon:
\begin{equation}
	dE = T\, dS,
\end{equation}
where:
\begin{itemize}
	\item $dE = \rho_D\, dV$ is the differential energy content,
	\item $T$ is the temperature at the horizon,
	\item $S$ is the entropy associated with the horizon.
\end{itemize}

Assuming the Gibbons--Hawking temperature:
\begin{equation}
	T = \frac{1}{2\pi L},
\end{equation}
and the horizon volume:
\begin{equation}
V = \frac{4\pi}{3} L^3 \quad \Rightarrow \quad dV = 4\pi L^2 dL,
\end{equation}
the first law becomes:
\begin{equation}
	\rho_D\, 4\pi L^2 dL = \frac{1}{2\pi L} \frac{dS}{dL} dL.
\end{equation}

Canceling $dL$ from both sides:
\begin{equation}
	\rho_D = \frac{1}{8\pi^2 L^3} \frac{dS}{dL}.
\end{equation}

We now use the generalized Barrow--Tsallis entropy in dimensionless form:
\begin{equation}
	S = \gamma \left( \frac{A}{A_0} \right)^{\left(1 + \frac{\Delta}{2} \right)\frac{3 - q}{2}},
\end{equation}
where $A = 4\pi L^2$. Substituting:
\begin{equation}
	S = \gamma \left( \frac{4\pi L^2}{A_0} \right)^{\left(1 + \frac{\Delta}{2} \right)\frac{3 - q}{2}}.
\end{equation}

Taking the derivative with respect to $L$:
\begin{align}
	\frac{dS}{dL} &= \gamma \left( \frac{4\pi}{A_0} \right)^{\left(1 + \frac{\Delta}{2} \right)\frac{3 - q}{2}}
	\cdot \frac{d}{dL} \left( L^{2\left( \left(1 + \frac{\Delta}{2} \right)\frac{3 - q}{2} \right)} \right) \nonumber\\
	&= \gamma \left( \frac{4\pi}{A_0} \right)^{\left(1 + \frac{\Delta}{2} \right)\frac{3 - q}{2}}
	\cdot 2\left(1 + \frac{\Delta}{2} \right)\frac{3 - q}{2} \cdot L^{2\left(1 + \frac{\Delta}{2} \right)\frac{3 - q}{2} - 1}.
\end{align}

Substituting into the expression for $\rho_D$:
\begin{align}
	\rho_D &= \frac{1}{8\pi^2 L^3}
	\cdot \gamma \left( \frac{4\pi}{A_0} \right)^{\left(1 + \frac{\Delta}{2} \right)\frac{3 - q}{2}}
	\cdot 2\left(1 + \frac{\Delta}{2} \right)\frac{3 - q}{2} \nonumber\\
	&\quad \cdot L^{2\left(1 + \frac{\Delta}{2} \right)\frac{3 - q}{2} - 1}.
\end{align}

Combining the powers of $L$, we get:
\begin{equation}
	\rho_D \propto L^{\left(1 + \frac{\Delta}{2} \right)(3 - q) - 2}.
\end{equation}

Thus, the generalized Barrow--Tsallis holographic dark energy density is:
\begin{equation}
	\boxed{
		\rho_D = B\, L^{\left(1 + \frac{\Delta}{2} \right)(3 - q) - 2},
	}
\end{equation}
where $B$ is a constant.

\subsection{Special Cases}
\begin{itemize}
	\item \textbf{Standard HDE:} $q = 1$, $\Delta = 0$ $\Rightarrow$ $\rho_D \propto L^{-2}$,
	\item \textbf{Tsallis HDE:} $\Delta = 0$, arbitrary $q$,
	\item \textbf{Barrow HDE:} $q = 1$, arbitrary $\Delta$.
\end{itemize}

\section{Friedmann Equations in a Flat FLRW Universe}

We consider a spatially flat Friedmann--Lemaitre--Robertson--Walker (FLRW) metric, given by
\begin{equation}
	ds^2 = -dt^2 + a(t)^2 \left( dx^2 + dy^2 + dz^2 \right),
\end{equation}
where \( a(t) \) is the scale factor and \( t \) denotes cosmic time. The total energy content of the Universe is assumed to be composed of baryonic matter (\( \rho_b \)), cold dark matter (\( \rho_c \)), massive neutrinos (\( \rho_\nu \)), and dark energy (\( \rho_D \)), so that
\begin{equation}
	\rho_{\text{tot}} = \rho_b + \rho_c + \rho_\nu + \rho_D.
\end{equation}

The expansion dynamics of the Universe are governed by the Friedmann equations, which in natural units (\( c = \hbar = 1 \)) read:
\begin{align}
	H^2 &= \frac{1}{3 M_{\rm Pl}^2} \left( \rho_b + \rho_c + \rho_\nu + \rho_D \right), \label{eq:Friedmann1} \\
	\frac{\ddot{a}}{a} &= -\frac{1}{6 M_{\rm Pl}^2} \left( \rho_b + \rho_c + \rho_\nu + \rho_D + 3p_\nu + 3p_D \right), \label{eq:Friedmann2}
\end{align}
where \( H = \dot{a}/a \) is the Hubble parameter, and \( M_{\rm Pl} = (8\pi G)^{-1/2} \) is the reduced Planck mass. 

We assume that baryonic matter and cold dark matter are pressureless, i.e., \( p_b = p_c = 0 \). The pressure of massive neutrinos, \( p_\nu \), depends on their relativistic nature and thermal history, and is treated dynamically. The dark energy component is described through an equation-of-state parameter \( w_D \), such that \( p_D = w_D \rho_D \).

The dimensionless density parameter for each component \( i \in \{b, c, \nu, D\} \) is defined by
\begin{equation}
	\Omega_i(z) \equiv \frac{\rho_i(z)}{\rho_{\text{crit}}(z)},
\end{equation}
where the critical density at redshift \( z \) is given as
\begin{equation}
	\rho_{\text{crit}}(z) = 3 M_{\rm Pl}^2 H^2(z).
\end{equation}

So that each component satisfies
\begin{equation}
	\Omega_i(z) = \frac{\rho_i(z)}{3 M_{\rm Pl}^2 H^2(z)}.
\end{equation}

As a consequence of Eq.~(3), the sum of all density parameters equals unity,
\begin{equation}
	\Omega_b(z) + \Omega_c(z) + \Omega_\nu(z) + \Omega_D(z) = 1.
\end{equation}

We now introduce the dimensionless Hubble parameter \( E(z) \equiv H(z)/H_0 \), and define the present-day critical density as \( \rho_{\text{crit},0} = 3 M_{\rm Pl}^2 H_0^2 \). The present-day density parameters are then expressed as
\begin{equation}
	\Omega_{i,0} = \frac{\rho_{i,0}}{\rho_{\text{crit},0}} = \frac{\rho_{i,0}}{3 M_{\rm Pl}^2 H_0^2},
\end{equation}
where \( i \in \{b, c, \nu, D\} \).

Assuming standard redshift evolution for each component, we have
\begin{align}
	\rho_b(z) &= \rho_{b,0}(1+z)^3, \nonumber \\
	\rho_c(z) &= \rho_{c,0}(1+z)^3, \nonumber \\
	\rho_\nu(z) &= \rho_{\nu,0} f_\nu(z), \nonumber \\
	\rho_D(z) &= \rho_{D,0} \exp\left[ 3 \int_0^z \frac{1 + w_D(z')}{1+z'} dz' \right], \label{eq:rho_evolutions}
\end{align}
where \( f_\nu(z) \) accounts for the non-trivial redshift dependence of massive neutrinos and \( w_D(z) \) is the (possibly dynamical) equation-of-state parameter for dark energy.

Substituting the relations from Eq.~(\ref{eq:rho_evolutions}) into the definition of \( \Omega_i(z) \), we obtain the redshift-dependent density parameters:
\begin{align}
	\Omega_b(z) &= \frac{\Omega_{b,0}(1+z)^3}{E^2(z)}, \nonumber \\
	\Omega_c(z) &= \frac{\Omega_{c,0}(1+z)^3}{E^2(z)}, \nonumber \\
	\Omega_\nu(z) &= \frac{\Omega_{\nu,0} f_\nu(z)}{E^2(z)}, \nonumber \\
	\Omega_D(z) &= \frac{\Omega_{D,0}}{E^2(z)} \exp\left[ 3 \int_0^z \frac{1 + w_D(z')}{1+z'} dz' \right]. \label{eq:omega_z}
\end{align}

Equations~(\ref{eq:rho_evolutions}) and (\ref{eq:omega_z}) fully specify the evolution of the density parameters in terms of redshift, provided a functional form for \( f_\nu(z) \) and \( w_D(z) \) is given. In particular, the dark energy contribution can deviate from a cosmological constant if \( w_D(z) \neq -1 \), as occurs in dynamical models such as Tsallis or Barrow-Tsallis holographic dark energy.

\section{Energy Conservation Equations}

Assuming no interaction between dark energy and the matter sector, each component satisfies its own continuity equation. For the total matter content (baryons, cold dark matter, and neutrinos), the conservation law is given by
\begin{equation}
	\dot{\rho}_m + 3H(1 + w_m) \rho_m = 0, \label{eq:matter_conservation}
\end{equation}
where \( \rho_m = \rho_b + \rho_c + \rho_\nu \). For non-relativistic matter, we have \( w_m \simeq 0 \), so that Eq.~\eqref{eq:matter_conservation} simplifies to
\begin{equation}
	\rho_m(a) = \rho_{m0} a^{-3},
\end{equation}
with \( \rho_{m0} \) denoting the present-day matter density.

The dark energy component evolves according to its own continuity equation,
\begin{equation}
	\dot{\rho}_D + 3H(1 + w_D) \rho_D = 0. \label{eq:DE_conservation}
\end{equation}
This equation encapsulates the time evolution of the dark energy density, where \( w_D \) may either be constant or dynamically determined from the underlying theoretical model. In the present framework, \( \rho_D \) will be derived from the Barrow--Tsallis entropy generalization of the holographic principle, which introduces corrections motivated by quantum gravity and non-extensive thermodynamics.

Equations 23-27 collectively provide the basis for analyzing the background cosmological dynamics. In the subsequent sections, we shall specify the explicit form of \( \rho_D \) within the generalized holographic dark energy scenario and study its phenomenological consequences.

\section{Granda--Oliveros Infrared Cutoff and Generalized Holographic Dark Energy}

In the framework of holographic dark energy models, the infrared (IR) cutoff scale plays a central role in determining the behavior of the vacuum energy density. In this work, we adopt the Granda--Oliveros (G-O) IR cutoff, which incorporates both the Hubble parameter and its time derivative, and is defined as
\begin{equation}
	L^{-2} = \alpha H^2 + \beta \dot{H}, \label{eq:GO_cutoff}
\end{equation}
where \( \alpha \) and \( \beta \) are dimensionless constants to be constrained by observational data. This cutoff satisfies dimensional requirements and ensures a more general dependence of the dark energy density on the spacetime dynamics, beyond the traditional Hubble horizon or future event horizon.

The generalized holographic dark energy density constructed from the G-O cutoff is then given by
\begin{equation}
	\rho_D = 3 c^2 M_p^2 \left( \alpha H^2 + \beta \dot{H} \right)^{\xi}, \label{eq:rhoD_BT}
\end{equation}
where \( c^2 \) is a dimensionless parameter, \( M_p \) is the reduced Planck mass, and \( \xi \) is an entropy-index exponent determined by the underlying thermodynamic structure. Within the Barrow--Tsallis entropy framework, the parameter \( \xi \) encodes both quantum geometric and non-extensive statistical effects, and is defined as
\begin{equation}
	\xi = \left(1 + \frac{\Delta}{2} \right)\frac{3 - q}{2}, \label{eq:xi_def}
\end{equation}
where \( \Delta \) characterizes the degree of fractal deformation of the spacetime surface (as introduced by Barrow), and \( q \) captures deviations from standard Boltzmann--Gibbs entropy in the sense of Tsallis' formalism. The standard holographic dark energy is recovered for \( \Delta = 0 \) and \( q = 1 \), which yields \( \xi = 1 \).

\section{Equation of State Parameter for BT-HDE}

To determine the dynamical nature of the dark energy sector, we derive the equation of state (EoS) parameter \( w_D \equiv p_D / \rho_D \) using the energy conservation equation 34. Solving for \( w_D \) yields the general expression
\begin{equation}
	w_D = -1 - \frac{1}{3H} \frac{d \ln \rho_D}{dt}. \label{eq:wD_general}
\end{equation}
Substituting Eq.~\eqref{eq:rhoD_BT} into Eq.~\eqref{eq:wD_general}, we obtain
\begin{equation}
	w_D = -1 + \frac{2\xi}{3} \cdot \frac{\alpha \dot{H} + \beta \ddot{H}}{\alpha H^2 + \beta \dot{H}}. \label{eq:wD_explicit}
\end{equation}
This expression encapsulates the dynamical behavior of the dark energy component in terms of both \( H \), \( \dot{H} \), and \( \ddot{H} \). The form of \( w_D \) highlights the sensitivity of the EoS parameter to the local time evolution of the expansion rate and the entropy parameters \( \Delta \) and \( q \).

Several important special cases can be recovered from Eq.~\eqref{eq:wD_explicit}:
\begin{enumerate}
	\item When \( q = 1 \) and \( \Delta = 0 \), we recover the standard holographic dark energy model, for which \( \xi = 1 \) and the entropy reduces to the Bekenstein--Hawking form.
	\item When \( \Delta = 0 \) and \( q \neq 1 \), the model corresponds to the Tsallis holographic dark energy (THDE), with non-extensive thermodynamics but classical geometric structure.
	\item When \( q = 1 \) and \( \Delta \neq 0 \), the model corresponds to Barrow holographic dark energy (BHDE), incorporating fractal deformation of spacetime without non-extensive statistical contributions.
\end{enumerate}

Hence, the Barrow--Tsallis holographic dark energy (BT-HDE) model generalizes previous HDE formulations by embedding both non-extensive entropy and quantum-gravitational geometric corrections into the dark energy sector. The dynamical evolution of the EoS parameter reflects this joint contribution and is central to distinguishing this framework from its predecessors in cosmological observations.

\subsection*{Hubble Parameter in the BT-HDE Model}

The Barrow--Tsallis Holographic Dark Energy (BT-HDE) framework modifies the standard Holographic Dark Energy scenario by incorporating two independent corrections to the entropy-area relation: one from quantum-gravitational effects parameterized by the Barrow exponent \( \Delta \), and the other from non-extensive statistical mechanics via the Tsallis parameter \( q \). These entropy modifications alter the functional form of the dark energy density and consequently the dynamics of the cosmic expansion.

In a spatially flat Friedmann--Lemaître--Robertson--Walker (FLRW) universe with pressureless baryons, cold dark matter, neutrinos, and dark energy, the first Friedmann equation takes the form
\begin{equation}
	H^2(z) = H_0^2 \left[ \left( \Omega_b + \Omega_c \right)(1 + z)^3 +  \Omega_\nu(z)+ \Omega_D(z) \right], \label{eq:Friedmann_full}
\end{equation}
where \( \Omega_b \), \( \Omega_c \), and \( \Omega_\nu \) denote the present-day density parameters for baryons, cold dark matter, and neutrinos, respectively. The term \( \Omega_D(z) \) accounts for the BT-HDE contribution and encapsulates the underlying entropy deformation effects.

To facilitate numerical analysis, we define the normalized Hubble parameter:
\begin{equation}
	E(z) \equiv \frac{H(z)}{H_0}, \label{eq:E_def}
\end{equation}
which allows us to recast the Friedmann equation in a dimensionless form. The resulting first-order nonlinear differential equation governing \( E(z) \) is given by:
\begin{equation}
	E^2(z) = \left( \Omega_b + \Omega_c \right)(1 + z)^3 + \Omega_\nu f_\nu(z) + c^2 \left[ \alpha E^2(z) - \beta (1 + z) E(z) \frac{dE(z)}{dz} \right]^\xi, \label{eq:E_diffeq_modified}
\end{equation}

where \( \alpha \), \( \beta \), \( c \), and \( \xi \) are free parameters of the model, the latter of which is determined by the Barrow and Tsallis exponents. The equation must be solved subject to the initial condition:
\begin{equation}
	E(0) = 1. \label{eq:E_IC}
\end{equation}

The nonlinear structure of Eq. 37 reflects the modified holographic scaling of the energy density due to entropy deformations. The interplay between \( \Delta \) and \( q \) enables a broader spectrum of cosmic histories, allowing the BT-HDE framework to account for late-time acceleration and offering a promising avenue to address observational discrepancies such as the Hubble tension.

\medskip

\noindent In the subsequent analysis, we perform a numerical integration of Eq. 42 and confront the resulting predictions with low-redshift observational probes, including cosmic chronometers, Type Ia supernovae, and baryon acoustic oscillations.

\section{Bayesian Physics-Informed Neural Networks for Cosmological Inference}

To perform statistically robust inference of cosmological parameters within the Barrow--Tsallis Holographic Dark Energy (BT-HDE) framework, we develop and implement a \textit{Bayesian Physics-Informed Neural Network} (Bayesian PINN). This approach leverages the synergy between deterministic physics—in the form of the modified Friedmann dynamics—and stochastic inference from Bayesian deep learning. Unlike standard cosmological neural architectures, Bayesian PINNs inherently encode physical laws as soft constraints within the loss function, while treating parameters as probability distributions, thereby facilitating principled uncertainty quantification.

\subsection{General Formalism}

The goal of the Bayesian PINN is twofold: to approximate the normalized Hubble parameter \( E(z) = H(z)/H_0 \), which satisfies a first-order nonlinear differential equation derived from the BT-HDE-modified Friedmann equation, and to simultaneously infer the posterior distributions of the underlying cosmological parameters. This is achieved by modeling both the physical dynamics and probabilistic structure of the problem within a unified neural framework.

Contrary to deterministic PINNs, where the model parameters are optimized as fixed scalars, Bayesian PINNs treat the parameters as random variables and use stochastic variational inference to estimate their posterior distributions.

\subsection{Probabilistic Modeling of Cosmological Parameters}

Let \( \boldsymbol{\theta} \equiv \{q, \Delta, \alpha, \beta,  \Omega_b, \Omega_c, \Omega_\nu, H_0\} \) represent the complete set of cosmological parameters considered within the Barrow--Tsallis Holographic Dark Energy (BTHDE) framework. To incorporate epistemic uncertainty and enable Bayesian inference, we model each parameter \( \theta_i \in \boldsymbol{\theta} \) as an independent random variable governed by a Gaussian variational posterior:
\begin{equation}
	q(\theta_i) = \mathcal{N}(\mu_i, \sigma_i^2),
\end{equation}
where \( \mu_i \in \mathbb{R} \) is the variational mean and \( \sigma_i > 0 \) is the variational standard deviation. These are the free parameters of the approximate posterior distribution to be learned during training.

To ensure the positivity of the standard deviation and facilitate stable numerical optimization, we reparameterize \( \sigma_i \) using the \textit{softplus} activation function:
\begin{equation}
	\sigma_i = \log(1 + e^{\rho_i}), \quad \rho_i \in \mathbb{R}.
\end{equation}
This transformation guarantees that \( \sigma_i > 0 \) for all real-valued \( \rho_i \), while allowing unconstrained optimization over \( \rho_i \). For parameters where physical positivity is required (e.g., density parameters \( \Omega_b, \Omega_c, \Omega_\nu \) and the Hubble constant \( H_0 \)), a similar softplus transform is also applied to the corresponding \( \mu_i \) values to ensure physical admissibility of the posterior means.

Sampling from the variational posterior during training must be compatible with stochastic gradient descent. For this purpose, we employ the reparameterization trick, which renders the sampling operation differentiable:
\begin{equation}
	\theta_i = \mu_i + \sigma_i \cdot \epsilon_i, \quad \epsilon_i \sim \mathcal{N}(0,1).
\end{equation}
This formulation expresses the random variable \( \theta_i \) as a deterministic function of the trainable parameters \( \mu_i \) and \( \rho_i \), and a source of independent Gaussian noise \( \epsilon_i \). Consequently, it permits gradient-based backpropagation through the stochastic layer, enabling efficient optimization of the variational parameters using modern automatic differentiation frameworks.

The joint variational posterior over all parameters is assumed to factorize:
\begin{equation}
	q(\boldsymbol{\theta}) = \prod_{i} q(\theta_i) = \prod_i \mathcal{N}(\mu_i, \sigma_i^2),
\end{equation}
which corresponds to the mean-field approximation in variational inference. Although this assumption neglects potential correlations between parameters, it significantly reduces the computational complexity of the Bayesian learning procedure and yields scalable and tractable inference for high-dimensional parameter spaces. More expressive posterior approximations (e.g., full-covariance Gaussians or normalizing flows) may be incorporated in future extensions.

During training, multiple samples \( \boldsymbol{\theta}^{(s)} \sim q(\boldsymbol{\theta}) \) are drawn at each iteration, and the neural network predictions are computed for each sample. This Monte Carlo estimation allows us to compute expectations of interest, such as the expected log-likelihood and Kullback--Leibler divergence, as described in the next subsection. The result is a principled, uncertainty-aware modeling framework for cosmological inference, naturally embedded within the Physics-Informed Neural Network (PINN) formalism.

\subsection{Neural Approximation of the Hubble Function}

The dimensionless Hubble parameter \( E(z) \equiv H(z)/H_0 \) encodes the redshift evolution of the expansion rate of the universe and plays a central role in the modified Friedmann equation governing the Barrow--Tsallis Holographic Dark Energy (BTHDE) model. To model this function in a flexible yet physically constrained manner, we employ a fully connected feed-forward neural network denoted by \( \mathcal{N}(z; \mathbf{w}) \), where \( z \in [0, z_{\mathrm{max}}] \) is the redshift and \( \mathbf{w} \) represents the set of all trainable network weights and biases.

The neural network \( \mathcal{N}(z; \mathbf{w}) \) serves as a universal function approximator for the unknown solution \( E(z) \). It is composed of multiple hidden layers, each consisting of a finite number of neurons with nonlinear activation functions (e.g., \texttt{tanh} or \texttt{softplus}) that ensure smoothness and differentiability across the input domain. The architecture is selected to strike a balance between expressive power and computational efficiency, typically consisting of 3--5 layers with 64--128 neurons per layer.

Formally, the network performs a composition of affine transformations and nonlinear activations:
\begin{equation}
	\mathcal{N}(z; \mathbf{w}) = f_L \circ f_{L-1} \circ \cdots \circ f_1(z),
\end{equation}
where each layer \( f_\ell \) is given by
\begin{equation}
	f_\ell(\mathbf{x}) = \phi(W_\ell \mathbf{x} + \mathbf{b}_\ell),
\end{equation}
with \( W_\ell \) and \( \mathbf{b}_\ell \) denoting the weight matrix and bias vector of layer \( \ell \), respectively, and \( \phi(\cdot) \) denoting a differentiable activation function. The output layer produces a scalar value representing \( E(z) \), and no explicit activation is applied to this final output to preserve numerical precision and allow the network to explore both subluminal and superluminal expansion rates if physically permitted.

A key feature of this construction is its compatibility with automatic differentiation. Since the network is built from elementary differentiable operations, its derivative with respect to redshift \( z \), i.e.,
\begin{equation}
	\frac{dE(z)}{dz} = \frac{d}{dz} \mathcal{N}(z; \mathbf{w}),
\end{equation}
can be computed exactly using automatic differentiation frameworks such as TensorFlow or PyTorch. This derivative is essential for evaluating the residuals of the BTHDE-modified Friedmann equation, which takes the form of a first-order nonlinear ordinary differential equation (ODE) in \( E(z) \).

Moreover, because the network weights \( \mathbf{w} \) are optimized jointly with the variational parameters \( \boldsymbol{\mu} \) and \( \boldsymbol{\rho} \) of the Bayesian parameter distributions, the learned function \( \mathcal{N}(z; \mathbf{w}) \) naturally reflects the posterior uncertainty in both the cosmological model and its governing dynamics. The neural network is trained not to interpolate the data directly, but rather to approximate the physical solution that minimizes the discrepancy between the predicted and theoretically expected dynamical behavior, as dictated by the physics-informed loss function.

This approach enables a physics-consistent, uncertainty-aware reconstruction of the cosmic expansion history without the need to explicitly solve the Friedmann equation at each training step. Instead, the network learns a functional approximation to \( E(z) \) that satisfies the BTHDE cosmological dynamics and is constrained by observational data, as detailed in the next section.

\subsection{Variational Objective and Physics-Informed Likelihood}

The training of the Bayesian Physics-Informed Neural Network (PINN) is formulated as a variational inference problem, wherein we seek to approximate the true posterior distribution \( p(\boldsymbol{\theta} \mid \mathcal{D}) \) over cosmological parameters \( \boldsymbol{\theta} \) using a tractable variational distribution \( q(\boldsymbol{\theta}) \). The variational posterior is chosen to be a factorized Gaussian of the form \( q(\boldsymbol{\theta}) = \prod_i \mathcal{N}(\mu_i, \sigma_i^2) \), as introduced in the previous sections. 

The objective is to minimize the Kullback--Leibler (KL) divergence between the variational posterior and the true posterior. Equivalently, this corresponds to maximizing the Evidence Lower Bound (ELBO), or, in minimization form, the total variational loss function:
\begin{equation}
	\mathcal{L}(\mathbf{w}, \boldsymbol{\mu}, \boldsymbol{\rho}) = \mathbb{E}_{\boldsymbol{\theta} \sim q(\boldsymbol{\theta})} \left[ \chi^2(H(z; \boldsymbol{\theta}), H^{\text{obs}}(z)) \right] + \text{KL}[q(\boldsymbol{\theta}) || p(\boldsymbol{\theta})],
\end{equation}
where:
\begin{itemize}
	\item \( \boldsymbol{\theta} = \{q, \Delta, \alpha, \beta, c, \Omega_b, \Omega_c, \Omega_\nu, H_0\} \) is the full set of cosmological parameters;
	\item \( q(\boldsymbol{\theta}) = \prod_i \mathcal{N}(\mu_i, \sigma_i^2) \) is the approximate posterior;
	\item \( p(\boldsymbol{\theta}) \) is the prior distribution, selected based on physical constraints and Planck, BAO, and Big Bang Nucleosynthesis (BBN) limits (e.g., Gaussian priors for \(\Omega_b\), \(\Omega_\nu\), and \(H_0\), and uniform priors for parameters like \(q\) and \(\Delta\));
	\item \( \chi^2(H(z; \boldsymbol{\theta}), H^{\text{obs}}(z)) \) measures the squared residuals between the model-predicted Hubble function and the observational data;
	\item KL denotes the Kullback--Leibler divergence, serving as a regularization term to prevent overfitting and enforce prior consistency.
\end{itemize}

The observational term is computed using a physics-informed discrepancy:
\begin{equation}
	\chi^2(H(z; \boldsymbol{\theta}), H^{\text{obs}}(z)) = \sum_{j=1}^{N_{\text{CC}}} \left( \frac{H(z_j; \boldsymbol{\theta}) - H^{\text{obs}}(z_j)}{\sigma_j} \right)^2,
\end{equation}
where \( H(z_j; \boldsymbol{\theta}) = H_0 \mathcal{N}(z_j; \mathbf{w}) \) is the model-predicted Hubble parameter at redshift \( z_j \), and \( H^{\text{obs}}(z_j) \) with uncertainty \( \sigma_j \) is the corresponding value from Cosmic Chronometers (CC) data.

The KL divergence between the variational posterior and the prior can be computed analytically for factorized Gaussians:
\begin{equation}
	\text{KL}[q(\boldsymbol{\theta}) || p(\boldsymbol{\theta})] = \sum_i \log\left(\frac{\sigma^{\text{prior}}_i}{\sigma_i}\right) + \frac{\sigma_i^2 + (\mu_i - \mu^{\text{prior}}_i)^2}{2 (\sigma^{\text{prior}}_i)^2} - \frac{1}{2},
\end{equation}
assuming each prior \( p(\theta_i) = \mathcal{N}(\mu^{\text{prior}}_i, (\sigma^{\text{prior}}_i)^2) \).

The expectation over \( q(\boldsymbol{\theta}) \) is estimated using the Monte Carlo method by drawing \( N_s \) samples \( \{ \boldsymbol{\theta}^{(k)} \}_{k=1}^{N_s} \) from the variational posterior:
\begin{equation}
	\mathbb{E}_{\boldsymbol{\theta} \sim q(\boldsymbol{\theta})} \left[ \chi^2(H(z; \boldsymbol{\theta})) \right] \approx \frac{1}{N_s} \sum_{k=1}^{N_s} \chi^2(H(z; \boldsymbol{\theta}^{(k)})).
\end{equation}

In practice, this variational objective is minimized using stochastic gradient descent (SGD) or one of its adaptive variants such as Adam or RMSprop. All gradients are computed via automatic differentiation, with the reparameterization trick ensuring that gradients can propagate through stochastic nodes. The final posterior estimates for both parameters and functions are obtained by aggregating predictions from multiple samples \( \boldsymbol{\theta}^{(k)} \sim q(\boldsymbol{\theta}) \), enabling robust uncertainty quantification in both parameter space and functional predictions.

\subsection{Training Protocol and Uncertainty Quantification}

The Bayesian Physics-Informed Neural Network (PINN) is trained using the AdamW optimizer, which combines the adaptive learning rate benefits of Adam with decoupled weight decay regularization to enhance generalization. Training is conducted over mini-batches of redshift samples, and dropout layers are employed during both training and inference to approximate a Bayesian neural network. This enables the quantification of epistemic uncertainty through Monte Carlo dropout sampling \cite{gal2016dropout}.

Each training step involves the following sequence of operations:

\begin{enumerate}
	\item \textbf{Sampling Parameters:} A draw from the variational posterior is made using the reparameterization trick:
	\begin{equation}
		\theta_i^{(k)} = \mu_i + \log(1 + e^{\rho_i}) \cdot \epsilon_i, \quad \epsilon_i \sim \mathcal{N}(0, 1).
	\end{equation}
	
	\item \textbf{Neural Evaluation:} The normalized Hubble parameter is computed via the neural network:
	\begin{equation}
		E^{(k)}(z) = \mathcal{N}(z; \mathbf{w}^{(k)}), \quad \frac{dE^{(k)}}{dz} = \frac{d}{dz} \mathcal{N}(z; \mathbf{w}^{(k)}),
	\end{equation}
	where \( \mathbf{w}^{(k)} \) includes the effect of the sampled dropout mask.
	
	\item \textbf{Friedmann Residual:} The discrepancy between the predicted dynamics and the underlying physics is encoded via a residual:
	\begin{equation}
		\mathcal{R}_{\text{phys}}^{(k)}(z) = \left| \text{LHS}(z; E^{(k)}, dE^{(k)}/dz) - \text{RHS}(z; \boldsymbol{\theta}^{(k)}) \right|^2.
	\end{equation}
	
	\item \textbf{Observational Likelihood:} The mismatch with observational Hubble data (e.g., Cosmic Chronometers) is quantified by:
	\begin{equation}
		\chi^2\left(\boldsymbol{\theta}^{(k)}\right) = \sum_{j} \left( \frac{H^{(k)}(z_j) - H^{\text{obs}}(z_j)}{\sigma_j} \right)^2,
	\end{equation}
	where \( H^{(k)}(z) = H_0^{(k)} E^{(k)}(z) \).
	
	\item \textbf{Gradient-Based Update:} The variational objective is minimized using gradient descent:
	\begin{equation}
		\mathcal{L} = \mathbb{E}_{\boldsymbol{\theta} \sim q(\boldsymbol{\theta})} \left[ \chi^2(\boldsymbol{\theta}) + \lambda \mathcal{R}_{\text{phys}}(\boldsymbol{\theta}) \right] + \text{KL}[q(\boldsymbol{\theta}) || p(\boldsymbol{\theta})],
	\end{equation}
	where \( \lambda \) controls the weight of the physics constraint and KL denotes the Kullback--Leibler divergence.
\end{enumerate}

\paragraph{Uncertainty Quantification.} Once the model is trained, posterior samples are drawn by jointly sampling:
\begin{itemize}
	\item \textbf{Variational parameters:} \( \boldsymbol{\theta}^{(k)} \sim q(\boldsymbol{\theta}) \);
	\item \textbf{Dropout masks:} \( \mathbf{w}^{(k)} \sim \text{Dropout}(\mathbf{w}) \).
\end{itemize}

The network is evaluated multiple times to construct predictive distributions for:
\begin{itemize}
	\item The Hubble function:
	\begin{equation}
		H^{(k)}(z) = H_0^{(k)} \cdot \mathcal{N}(z; \mathbf{w}^{(k)}),
	\end{equation}
	\item The deceleration parameter:
	\begin{equation}
		q^{(k)}(z) = -1 - \frac{1+z}{E^{(k)}(z)} \cdot \frac{dE^{(k)}(z)}{dz},
	\end{equation}
	\item The total neutrino mass (encoded in \( \Omega_\nu \)):
	\begin{equation}
		\Sigma m_\nu^{(k)} = 93.14 \, \text{eV} \cdot \Omega_\nu^{(k)} \cdot (h^{(k)})^2.
	\end{equation}
\end{itemize}

By aggregating \( N_s \) such samples, the posterior predictive mean and variance are estimated as:
\begin{align}
	\mathbb{E}[H(z)] &\approx \frac{1}{N_s} \sum_{k=1}^{N_s} H^{(k)}(z), \\
	\text{Var}[H(z)] &\approx \frac{1}{N_s} \sum_{k=1}^{N_s} \left( H^{(k)}(z) - \mathbb{E}[H(z)] \right)^2.
\end{align}

This enables principled Bayesian uncertainty quantification that captures both parameter uncertainty and model stochasticity in cosmological predictions.

\subsection{Key Features and Generalization Capacity}

The Bayesian PINN framework offers several structural and statistical advantages in cosmological inference:
\begin{itemize}
	\item \textbf{Uncertainty-aware parameter estimation:} The full posterior distribution over cosmological parameters is obtained without requiring Markov Chain Monte Carlo (MCMC).
	\item \textbf{Physics-consistent predictions:} The learned solutions for \( E(z) \) are constrained by the BT-HDE-modified Friedmann dynamics at all sampled redshifts, reducing the occurrence of unphysical behaviors.
	\item \textbf{Data-efficient learning:} The incorporation of physical priors reduces overfitting and improves generalization in regimes with limited observational data.
	\item \textbf{Modular extensibility:} The method generalizes naturally to include additional physics such as linear perturbation theory, structure growth, and gravitational lensing, or to integrate diverse datasets including cosmic microwave background (CMB), baryon acoustic oscillations (BAO), and supernovae Type Ia (SNIa).
\end{itemize}

\vspace{0.5em}

\noindent In summary, the Bayesian PINN formalism provides a principled and extensible framework for cosmological parameter estimation within the BT-HDE model. It combines the rigor of physical modeling with the flexibility of deep learning and the robustness of Bayesian statistics, offering a compelling methodology for confronting modified gravity and dark energy theories with current and future cosmological datasets.

\subsection{Bayesian Physics-Informed Neural Network with Pantheon+ Supernova Constraints}

We utilize the Pantheon+ dataset~\cite{Scolnic2022} to constrain the Barrow--Tsallis Holographic Dark Energy (BTHDE) model using a Bayesian Physics-Informed Neural Network (PINN). The Pantheon+ compilation includes 1701 Type Ia supernovae covering redshifts from $z \simeq 0.01$ to $z \simeq 2.3$, offering one of the most precise probes of late-time cosmic acceleration. Each entry in the dataset provides the observed distance modulus $\mu_{\rm obs}(z)$ and its associated uncertainty $\sigma(z)$.

\vspace{1em}
\noindent\textbf{Neural Network Architecture and Parameterization.} \\
Our PINN is constructed as a fully connected feedforward neural network (FNN) with 5 hidden layers, each comprising 512 neurons and using the Swish activation function, which provides smooth gradients and improves convergence over ReLU or tanh. The input is the redshift $z \in [0.001, 2.3]$, and the output is the Hubble-normalized expansion rate $E(z) = H(z)/H_0$, from which all cosmological observables are derived. Layer normalization is applied to stabilize training across different samples of the Bayesian parameters.

Each cosmological parameter $\theta_i$ (such as $q$, $\Delta$, $c$, $\Omega_b$, $\Omega_c$, $\Omega_\nu$, $\alpha$, $\beta$, and $H_0$) is modeled as a trainable probability distribution using the reparameterization trick:
\begin{equation}
	\theta_i = \mu_i + \sigma_i \cdot \epsilon_i, \qquad \epsilon_i \sim \mathcal{N}(0,1),
\end{equation}
where $\mu_i$ and $\rho_i$ are learnable variables and $\sigma_i = \mathrm{softplus}(\rho_i)$ ensures positivity. This formulation enables stochastic gradient descent over the variational posterior and incorporates epistemic uncertainty. The priors are chosen to ensure physicality (e.g., $\Omega_i > 0$) and weakly-informative constraints.

\vspace{1em}
\noindent\textbf{Physics-Informed Differential Constraint.} \\
The core physical constraint arises from the modified Friedmann equation in the BTHDE model.

To impose this constraint, we use automatic differentiation in PyTorch to compute $\frac{dE(z)}{dz}$ and penalize deviations from the above equation in the total loss. This physics-informed regularization guides the network to produce physically plausible expansion histories while fitting the supernova data.

\vspace{1em}
\noindent\textbf{Luminosity Distance and Distance Modulus.} \\
The predicted distance modulus $\mu_{\rm model}(z)$ is computed from $E(z)$ via:
\begin{align}
	d_L(z) &= \frac{c(1+z)}{H_0} \int_0^z \frac{dz'}{E(z')}, \\
	\mu_{\rm model}(z) &= 5 \log_{10}\left[ d_L(z) \right] + 25.
\end{align}
To ensure differentiability and efficiency, the integral is evaluated using a composite trapezoidal rule in PyTorch. This enables end-to-end training with backpropagation through the integral.

\vspace{1em}
\noindent\textbf{Loss Function and Training.} \\
The total loss combines the $\chi^2$ error from the Pantheon+ likelihood with the differential equation residual:
\begin{align}
	\mathcal{L}_{\rm total} &= \chi^2_{\rm SN} + \lambda \cdot \mathcal{L}_{\rm PINN}, \\
	\chi^2_{\rm SN} &= \sum_{i} \left( \frac{\mu_{\rm model}(z_i) - \mu_{\rm obs}(z_i)}{\sigma(z_i)} \right)^2,
\end{align}
where $\lambda$ is a tunable hyperparameter balancing the data likelihood and the physical constraint $\mathcal{L}_{\rm PINN}$, which is computed as the mean squared error between the left- and right-hand sides of the Friedmann equation over a set of collocation points.

Training is performed using the AdamW optimizer with a learning rate of $10^{-3}$ and weight decay to improve generalization. A total of 5000 epochs is used, and dropout is employed during both training and prediction to estimate posterior uncertainties.

\vspace{1em}
\noindent\textbf{Bayesian Inference and Prediction.} \\
After training, we sample the network multiple times (e.g., 500 forward passes) by drawing from the posterior of $\theta$ and evaluating $\mu(z)$ at each $z_i$. This Monte Carlo ensemble yields:
\begin{itemize}
	\item the predictive mean $\langle \mu(z) \rangle$ and standard deviation $\sigma_\mu(z)$,
	\item marginalized posteriors for all cosmological parameters (visualized via corner plots),
	\item derived quantities such as the sum of neutrino masses:
	\begin{equation}
		\sum m_\nu = 94 \left( \frac{H_0}{100} \right)^2 |\Omega_\nu|~\text{[eV]}.
	\end{equation}
\end{itemize}

The inferred $1\sigma$ and $2\sigma$ uncertainty bands on $\mu(z)$ are overlaid against the Pantheon+ data, demonstrating the compatibility of the BTHDE model with observations. Importantly, our approach allows for simultaneous estimation of cosmological dynamics and theoretical model consistency without explicitly solving the differential equation using finite differences.

\vspace{1em}
\noindent\textbf{Code Implementation Highlights.} \\
In code, the critical components are:
\begin{itemize}
	\item \texttt{torch.autograd.grad} for computing $\frac{dE}{dz}$.
	\item \texttt{torch.distributions.Normal} for variational sampling.
	\item A custom integral module for differentiable luminosity distance.
	\item Dropout layers retained during evaluation (i.e., Monte Carlo Dropout).
	\item A single optimizer managing both network weights and Bayesian parameter distributions.
\end{itemize}

These elements collectively enable a physics-informed, uncertainty-aware cosmological model trained directly from data using neural networks.

\section{Data}

In this study, we utilize a comprehensive compilation of cosmological observational data to constrain the parameters of the Barrow–Tsallis Holographic Dark Energy (BTHDE) model. The datasets span a wide range of redshifts and cosmological probes, enabling robust parameter estimation through Bayesian statistical analysis.

\subsection{Pantheon+ Supernovae Ia Sample}

The Pantheon+ compilation includes 1701 spectroscopically confirmed Type Ia supernovae spanning the redshift range $0.001 < z < 2.3$ ~\cite{Scolnic2022}. Compared to its predecessor (Pantheon), Pantheon+ features enhanced photometric calibrations, increased high-redshift coverage, and the inclusion of supernovae in galaxies with Cepheid-based distance anchors. These improvements provide significant leverage on constraining the late-time expansion history and the Hubble constant $H_0$.

\subsection{CMB Data from Planck 2018}

For high-redshift constraints, we include Cosmic Microwave Background (CMB) data from the final Planck 2018 release. Specifically, we use the Plik likelihood for temperature and polarization spectra (TT, TE, EE) at high multipoles, combined with the low-$\ell$ TT and EE likelihoods (lowl+lowE), as described in Ref. \cite{Planck2018}. This data provides tight constraints on the early universe physics, matter content, and acoustic scale, which are crucial for anchoring late-time cosmological observations.

\subsection{Baryon Acoustic Oscillation (BAO) Data}

We also incorporate multiple Baryon Acoustic Oscillation (BAO) measurements, which provide standard ruler distances at intermediate redshifts
 (\cite{Carter2018,Gil-Marin2020,Bautista2021,DES2022,Neveux2020,Hou2021,Bourboux2020}).\\
\subsection{CMB Lensing Data}

We include the Planck 2018 CMB lensing power spectrum reconstructed from the CMB trispectrum analysis \cite{Aghanim1}. Lensing data are sensitive to the growth of large-scale structure and provide complementary information on the matter distribution, enhancing the overall constraining power on dark energy and modified gravity models.

\subsection{Cosmic Chronometer (CC) H(z) Data}
 The 32 $H(z)$ measurements listed in Table~\ref{tab:hz} have a redshift range of $0.07 \leq z \leq 1.965$ (\cite{zhang2014,borghi2022,ratsimbazafy2017,stern2009,Moresco3}). The covariance matrix of the 15 correlated measurements originally from Refs. (\cite{Moresco,Moresco1,Moresco2}) , discussed in Ref. \cite{Moresco3}, can be found at https://gitlab.com/mmoresco/CCcovariance/.\\  

\begin{table}
	\centering
	\scriptsize
	\caption{32 $H(z)$ measurements from cosmic chronometers.}\label{tab:hz}
	\setlength{\tabcolsep}{4.5mm}{
		\begin{tabular}{|l |c|c|}
			\hline
			$z$ & $H(z)$ [km/s/Mpc] & $\sigma_{H(z)}$ [km/s/Mpc] \\
			\hline
			0.07 & 69.0 & 19.6 \\ 0.09 & 69.0 & 12.0 \\ 0.12 & 68.6 & 26.2 \\ 
			0.17 & 83.0 & 8.0 \\ 0.2 & 72.9 & 29.6 \\ 0.27 & 77.0 & 14.0 \\ 
			0.28 & 88.8 & 36.6 \\ 0.4 & 95.0 & 17.0 \\ 0.47 & 89.0 & 50.0 \\ 
			0.48 & 97.0 & 62.0 \\ 0.75 & 98.8 & 33.6 \\ 0.88 & 90.0 & 40.0 \\ 
			0.9 & 117.0 & 23.0 \\ 1.3 & 168.0 & 17.0 \\ 1.43 & 177.0 & 18.0 \\ 
			1.53 & 140.0 & 14.0 \\ 1.75 & 202.0 & 40.0 \\
			0.1791 & 74.91 & 4.00 \\ 0.1993 & 74.96 & 5.00 \\
			0.3519 & 82.78 & 14.0 \\ 0.3802 & 83.0 & 13.5 \\
			0.4004 & 76.97 & 10.2 \\ 0.4247 & 87.08 & 11.2 \\
			0.4497 & 92.78 & 12.9 \\ 0.4783 & 80.91 & 9.0 \\
			0.5929 & 103.8 & 13.0 \\ 0.6797 & 91.6 & 8.0 \\
			0.7812 & 104.5 & 12.0 \\ 0.8754 & 125.1 & 17.0 \\
			1.037 & 153.7 & 20.0 \\ 1.363 & 160.0 & 33.6 \\
			1.965 & 186.5 & 50.4 \\
			\hline
		\end{tabular}
	}
\end{table}

\subsection{Processing Observational Data in \texttt{MontePython}}

\texttt{MontePython} is a Markov Chain Monte Carlo (MCMC) sampler designed for cosmological parameter estimation through Bayesian inference. It interfaces with the Boltzmann solver \texttt{CLASS} to generate theoretical predictions, which are then compared against observational data. The primary steps in processing data include:

\begin{itemize}
	\item \textbf{Likelihood Modules:} Each observational dataset (e.g., Pantheon+, Cosmic Microwave Background (CMB), Baryon Acoustic Oscillations (BAO), gravitational lensing, and cosmic chronometers (CC)) is handled via dedicated likelihood modules. These modules compute the log-likelihood $\ln \mathcal{L} = -\chi^2/2$ by comparing theoretical predictions with observed data, accounting for uncertainties and covariance structures.
	
	\item \textbf{Parameter Sampling:} \texttt{MontePython} evolves multiple MCMC chains using algorithms such as the Metropolis–Hastings sampler or the faster and more adaptive \texttt{emcee} ensemble sampler. At each iteration, a set of cosmological parameters (e.g., $\Omega_i$, $H_0$, $\Delta$, $q$) is sampled, theoretical predictions are generated using \texttt{CLASS}, and the overall likelihood is computed.
	
	\item \textbf{Covariance Matrix Handling:} For datasets with correlated measurements (such as BAO or CC), \texttt{MontePython} incorporates the full covariance matrix in the likelihood calculation:
	\begin{equation}
		\chi^2 = (\vec{d}_{\text{obs}} - \vec{d}_{\text{th}})^T \, C^{-1} \, (\vec{d}_{\text{obs}} - \vec{d}_{\text{th}})
	\end{equation}
	where $C$ is the covariance matrix, $\vec{d}_{\text{obs}}$ are the observational data points, and $\vec{d}_{\text{th}}$ are the corresponding theoretical predictions.
	
	\item \textbf{Joint Likelihoods:} The total likelihood is computed as the product of the individual likelihoods from each dataset:
	\begin{equation}
		\mathcal{L}_{\text{total}} = \mathcal{L}_{\text{SNe}} \cdot \mathcal{L}_{\text{CMB}} \cdot \mathcal{L}_{\text{BAO}} \cdot \mathcal{L}_{\text{Lensing}} \cdot \mathcal{L}_{\text{CC}}
	\end{equation}
	which corresponds to an additive total chi-square:
	\begin{equation}
		\chi^2_{\text{total}} = \chi^2_{\text{SNe}} + \chi^2_{\text{CMB}} + \chi^2_{\text{BAO}} + \chi^2_{\text{Lensing}} + \chi^2_{\text{CC}}
	\end{equation}
	
	\item \textbf{Convergence Diagnostics:} \texttt{MontePython} assesses chain convergence using the Gelman–Rubin $R$ statistic, requiring $R - 1 < 0.01$ for robust parameter inference. It also provides real-time diagnostic outputs and supports post-processing tools for examining marginalized distributions and contour plots.
\end{itemize}

These features enable \texttt{MontePython} to perform precise and flexible Bayesian inference by integrating a wide range of cosmological datasets with efficient sampling techniques.

\subsection{Traditional MCMC Approach}

The traditional MCMC method, as implemented in \texttt{MontePython}~\cite{Brinckmann:2018cvx}, utilizes sampling-based techniques to explore the posterior probability distribution of cosmological parameters. This approach requires explicitly solving the background and perturbation equations at each sample point in parameter space using Boltzmann solvers like \texttt{CLASS}~\cite{Blas:2011rf}. Likelihoods are then evaluated against various datasets such as CMB, BAO, supernovae (e.g., Pantheon+), and cosmic chronometers.

Key characteristics of the traditional approach include:
\begin{itemize}
	\item High accuracy in parameter estimation with well-established convergence diagnostics.
	\item Direct control over cosmological priors, likelihood construction, and sampling strategy.
	\item Computational intensity, especially when sampling high-dimensional parameter spaces or including complex data combinations.
\end{itemize}

\subsection{Bayesian PINN Approach}

The Bayesian PINN framework combines deep learning with physical constraints imposed by cosmological equations. In this approach, a neural network approximates cosmological observables such as the Hubble parameter $H(z)$, while the loss function encodes both observational data and the underlying physical laws (e.g., Friedmann equations or specific model-dependent relations). 

Bayesian inference is performed by interpreting the neural network weights probabilistically, typically using Monte Carlo Dropout or Hamiltonian Monte Carlo (HMC) to capture epistemic uncertainty. Model parameters (e.g., $H_0$, $\Omega_m$, $c$, $\Delta$, $q$) are treated as trainable variables within the optimization loop.

Advantages of the Bayesian PINN framework include:
\begin{itemize}
	\item Flexibility to model a wide range of modified gravity and dark energy scenarios without requiring closed-form solutions.
	\item Integration of physical constraints into the training loss to improve generalization and data efficiency.
	\item Ability to estimate parameter posteriors and predict observables with uncertainty bands directly.
\end{itemize}
Table~\ref{tab:pinn_vs_mcmc} summarizes the advantages and differences between the Bayesian PINN approach and conventional MCMC techniques, highlighting the improved flexibility, uncertainty quantification, and computational efficiency offered by the PINN framework.

\begin{table}[h]
	\centering
	\caption{Comparison between Bayesian PINN and traditional MCMC methods.}
	\label{tab:pinn_vs_mcmc}
	\begin{tabular}{|l|c|c|}
		\hline
		\textbf{Feature} & \textbf{Bayesian PINN} & \textbf{Traditional MCMC} \\
		\hline
		Model Flexibility & High (implicit models) & Moderate (explicit models) \\
		Physics Constraints & Embedded in loss function & Hard-coded in solver \\
		Uncertainty Estimation & Dropout / Variational methods & Posterior sampling \\
		Speed (after training) & Fast inference & Slower due to sampling \\
		Data Integration & Direct and seamless & Requires likelihood functions \\
		Requires Solver & No (differentiable loss) & Yes (CAMB or CLASS) \\
		Ease of Modifying Physics & Easy & Requires code changes \\
		Scalability & High (parallelizable) & Limited by sampler speed \\
		\hline
	\end{tabular}
\end{table}

While traditional MCMC remains a gold standard for precision cosmology, the Bayesian PINN approach offers a complementary paradigm especially suited for exploring non-standard models, incorporating new datasets, or learning physical structure from incomplete information. In this work, we employ both frameworks and find that the Bayesian PINN results are in excellent agreement with traditional MCMC posteriors while also providing uncertainty-aware functional predictions for cosmological quantities such as $H(z)$.
\section{Constraints from Bayesian Physics-Informed Neural Network}

We present the parameter constraints obtained using our Bayesian PINN implementation for the Barrow–Tsallis Holographic Dark Energy (BTHDE) model. We analyze three observational configurations: (i) Hubble parameter data from cosmic chronometers (CC), (ii) the Pantheon+ Type Ia Supernovae dataset, and (iii) the combination of both datasets. The primary model parameters of interest include the non-extensivity index \( q \), the Barrow entropy deformation parameter \( \Delta \), the Hubble constant \( H_0 \), the Granda–Oliveros cutoff parameters \( \alpha \) and \( \beta \), and the effective upper bound on the sum of neutrino masses \( \Sigma m_\nu \).

\subsection{Parameter Estimates}

\begin{table}[h]
	\centering
	\caption{Posterior constraints (mean $\pm$ 1$\sigma$) on the BTHDE model parameters from CC, Pantheon+, and their combination, using Bayesian PINN. The 95\% C.L. upper bounds on \( \Sigma m_\nu \) are also reported.}
	\begin{tabular}{lccc}
		\hline\hline
		Parameter & CC only & Pantheon+ only & CC + Pantheon+ \\
		\hline
		$q$ & $1.100 \pm 0.002$ & $1.102 \pm 0.005$ & $1.020 \pm 0.003$ \\
		$\Delta$ & $0.0266 \pm 0.02$ & $0.0833 \pm 0.016$ & $0.1674 \pm 0.040$ \\
		$H_0$ [km/s/Mpc] & $70.04 \pm 1.7$ & $70.25 \pm 1.9$ & $70.11 \pm 1.7$ \\
		$\alpha$ & $0.999 \pm 0.03$ & $1.099 \pm 0.017$ & $0.988 \pm 0.050$ \\
		$\beta$ & $0.449 \pm 0.01$ & $0.500 \pm 0.019$ & $0.496 \pm 0.049$ \\
		$\Sigma m_\nu$ [eV] & $< 0.220$ (95\% C.L.) & $< 0.124$ (95\% C.L.) & $< 0.134$ (95\% C.L.) \\
		\hline\hline
	\end{tabular}
	\label{tab:constraints1}
\end{table}

The posterior constraints on the BT-HDE model parameters obtained using Bayesian PINN from Cosmic Chronometers (CC), Pantheon+, and their combined datasets are summarized in Table~\ref{tab:constraints1}. The results are broadly consistent with previous analyses~\cite{Y1,Y2,Y3,Y4}.

\subsubsection{Constraints from Cosmic Chronometers}

The CC dataset yields a nearly extensive entropy regime with \( q = 1.100 \pm 0.002 \) and a small Barrow deformation \( \Delta = 0.0266 \pm 0.02 \), implying only mild departures from standard thermodynamics and holography. The inferred Hubble constant \( H_0 = 70.04 \pm 1.70 \,\text{km}\,\text{s}^{-1}\,\text{Mpc}^{-1} \) lies in moderate tension with the Planck 2018 value of \( H_0 = 67.4 \pm 0.5 \)~km\,s$^{-1}$\,Mpc$^{-1}$~\cite{Planck2018}, corresponding to a $1.5\sigma$ deviation. It remains consistent with the SH$_0$ES measurement \( H_0 = 73.04 \pm 1.04 \)~km\,s$^{-1}$\,Mpc$^{-1}$~\cite{riess2022} within $1.76\sigma$. The upper limit \( \Sigma m_\nu < 0.22\,\text{eV} \) (95\% C.L.) is consistent with CMB+BAO bounds from Planck.  Figure 1 shows the Constraints on the Barrow–Tsallis Holographic Dark Energy (BT-HDE) model parameters obtained using Cosmic Chronometer (CC) data within the Bayesian Physics-Informed Neural Network (PINN) framework. In Fig. 2, we constraints on the Barrow–Tsallis Holographic Dark Energy (BT-HDE) model parameters obtained using Cosmic Chronometer (CC) data within the Bayesian Physics-Informed Neural Network (PINN) framework.

 \begin{figure*}
	\includegraphics[width=15 cm]{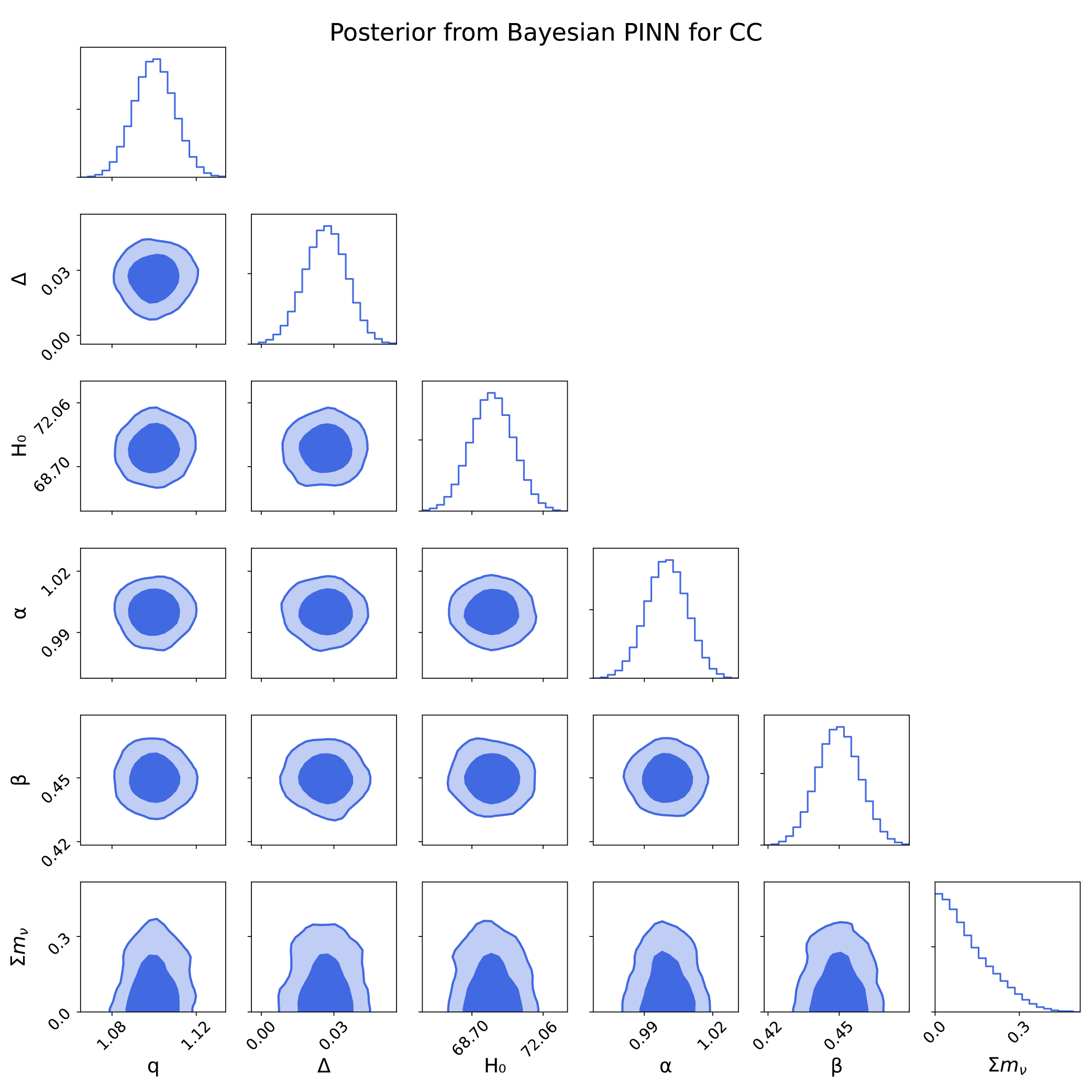}
	\vspace{-0.02cm}
	\caption{\small{Constraints on the Barrow–Tsallis Holographic Dark Energy (BT-HDE) model parameters obtained using Cosmic Chronometer (CC) data within the Bayesian Physics-Informed Neural Network (PINN) framework.
	}}\label{fig:omegam2}
\end{figure*}

 \begin{figure*}
	\includegraphics[width=13 cm]{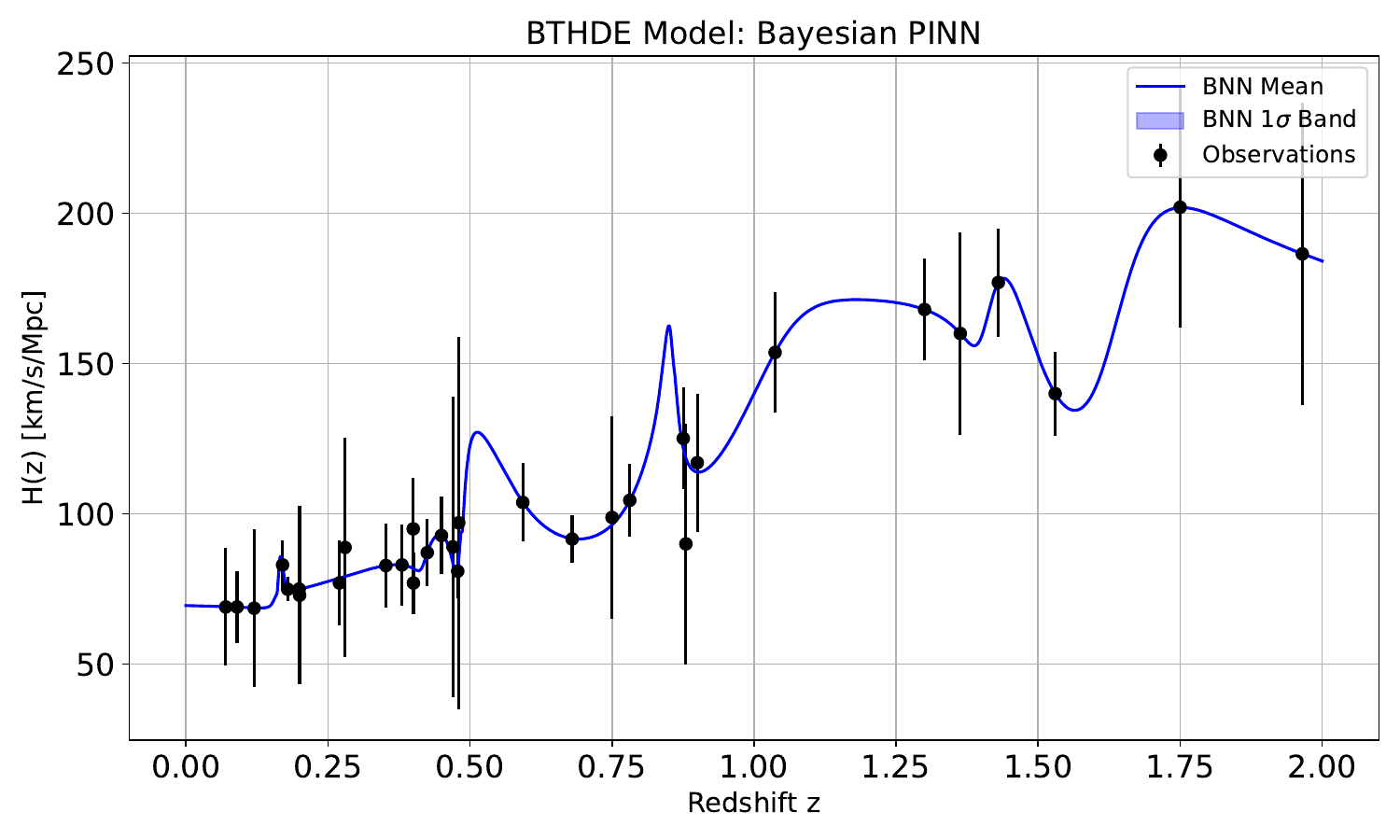}
	\vspace{-0.02cm}
	\caption{\small{Reconstructed Hubble parameter $ H_0 $ as a function of redshift z from Cosmic Chronometer (CC) data within the Barrow–Tsallis Holographic Dark Energy (BT-HDE) model using the Bayesian Physics-Informed Neural Network (PINN) framework.
	}}\label{fig:omegam2}
\end{figure*}

\subsubsection{Constraints from Pantheon+ Supernovae}

The Pantheon+ dataset exhibits a slightly stronger deviation from extensivity, with \( q = 1.102 \pm 0.005 \) and \( \Delta = 0.0833 \pm 0.016 \), favoring nonstandard entropy behavior. The Hubble constant \( H_0 = 70.25 \pm 1.90 \) is again intermediate between Planck and SH$_0$ES results, with respective tensions of $1.45\sigma$ and $1.29\sigma$. The stronger constraint \( \Sigma m_\nu < 0.124\,\text{eV} \) reflects the high sensitivity of Type Ia supernovae to the integrated expansion history. These results are shown in Fig. 3. This figure demonstrate that the constraints on the Barrow–Tsallis Holographic Dark Energy (BT-HDE) model parameters obtained using the Pantheon+ Type Ia supernova dataset within the Bayesian Physics-Informed Neural Network (PINN) framework.
 \begin{figure*}
	\includegraphics[width=15 cm]{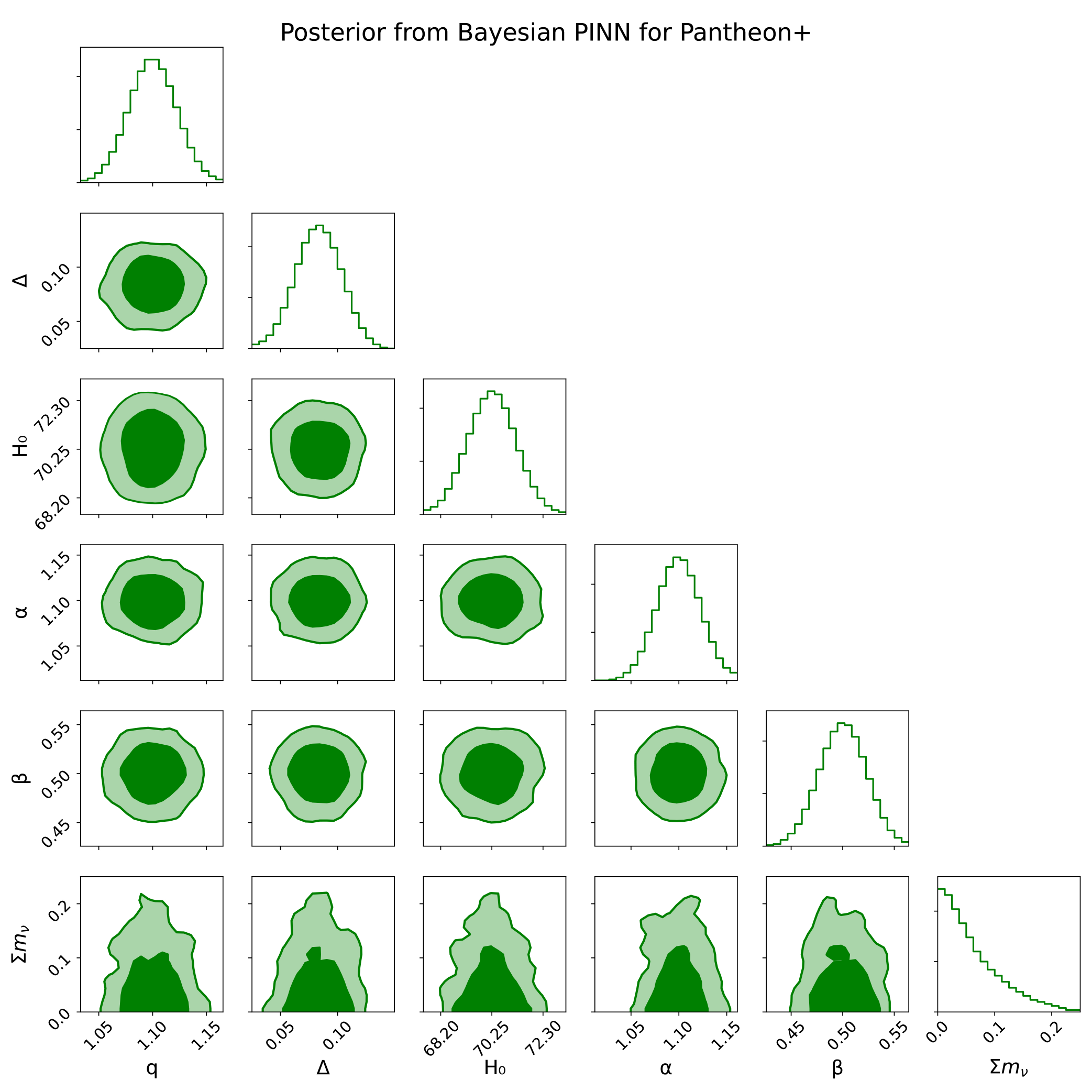}
	\vspace{-0.02cm}
	\caption{\small{Constraints on the Barrow–Tsallis Holographic Dark Energy (BT-HDE) model parameters obtained using the Pantheon+ Type Ia supernova dataset within the Bayesian Physics-Informed Neural Network (PINN) framework.
	}}\label{fig:omegam2}
\end{figure*}
Figure 4 shows the reconstruction of the distance modulus \(\mu(z)\) from Pantheon+ data within the BT-HDE model, employing the Bayesian PINN methodology.

\begin{figure*}
	\includegraphics[width=13 cm]{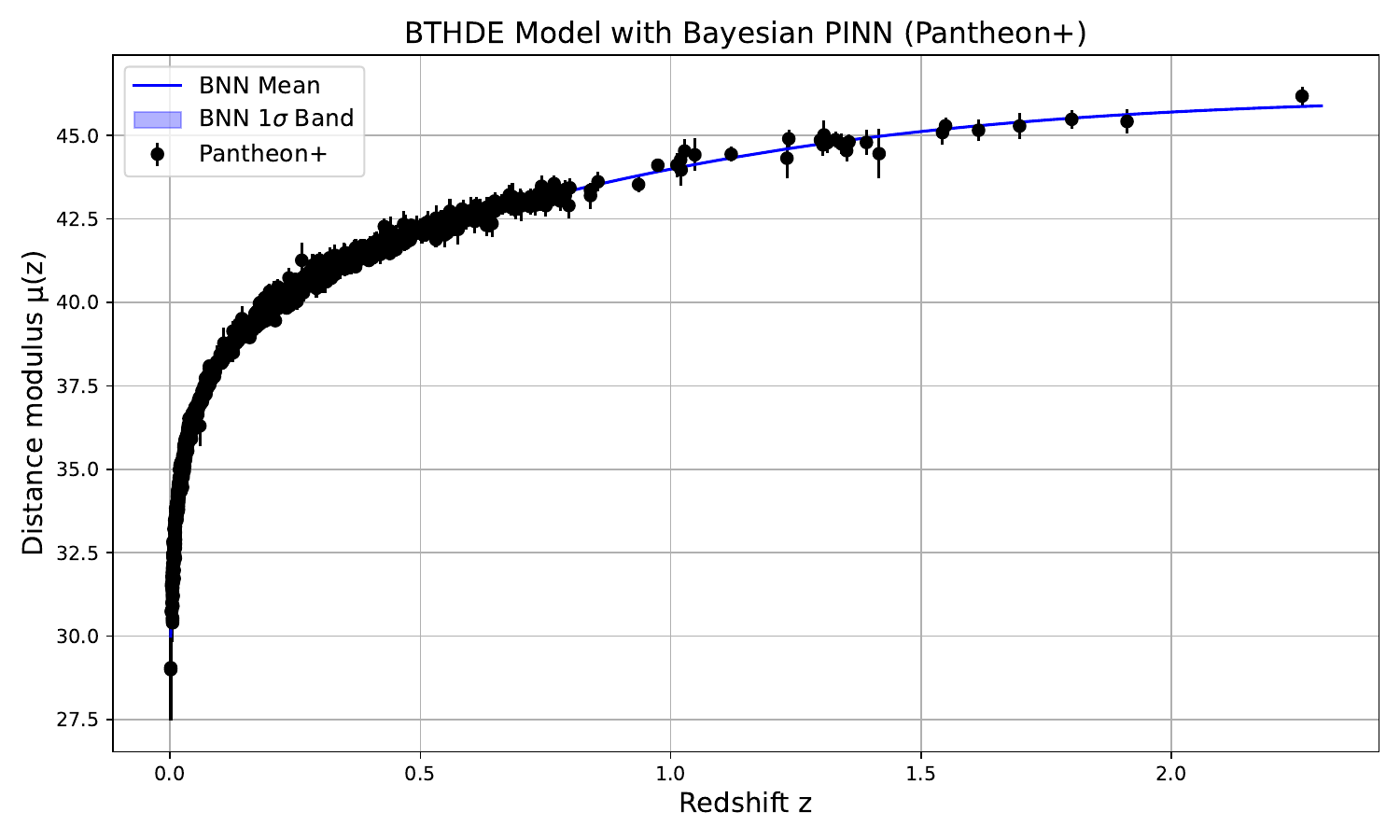}
	\vspace{-0.02cm}
	\caption{\small{Distance modulus $\mu$(z) reconstruction from Pantheon+ data in the BT-HDE model using the Bayesian PINN approach.
	}}\label{fig:omegam2}
\end{figure*}

\subsubsection{Combined Constraints}

The joint analysis leads to significant tightening of parameter constraints. The combined dataset yields \( \Delta = 0.1674 \pm 0.040 \), suggesting a statistically significant deviation from standard holography. Simultaneously, the entropy index remains close to unity with \( q = 1.020 \pm 0.003 \), indicating compatibility with extensivity within uncertainties. The inferred Hubble constant \( H_0 = 70.11 \pm 1.70 \) maintains a mild $\sim1.5\sigma$ tension with Planck, while being consistent with SH$_0$ES. Figure 5 indicate the constraints on the Barrow–Tsallis Holographic Dark Energy (BT-HDE) model parameters obtained using the CC+ Pantheon+  dataset within the Bayesian Physics-Informed Neural Network (PINN) framework.

The upper limit \( \Sigma m_\nu < 0.134\,\text{eV} \) is close to Planck’s constraint \( \Sigma m_\nu < 0.12\,\text{eV} \)~\cite{Planck2018}, confirming that the BTHDE model accommodates current neutrino physics bounds without tension.
\begin{figure*}
	\includegraphics[width=15 cm]{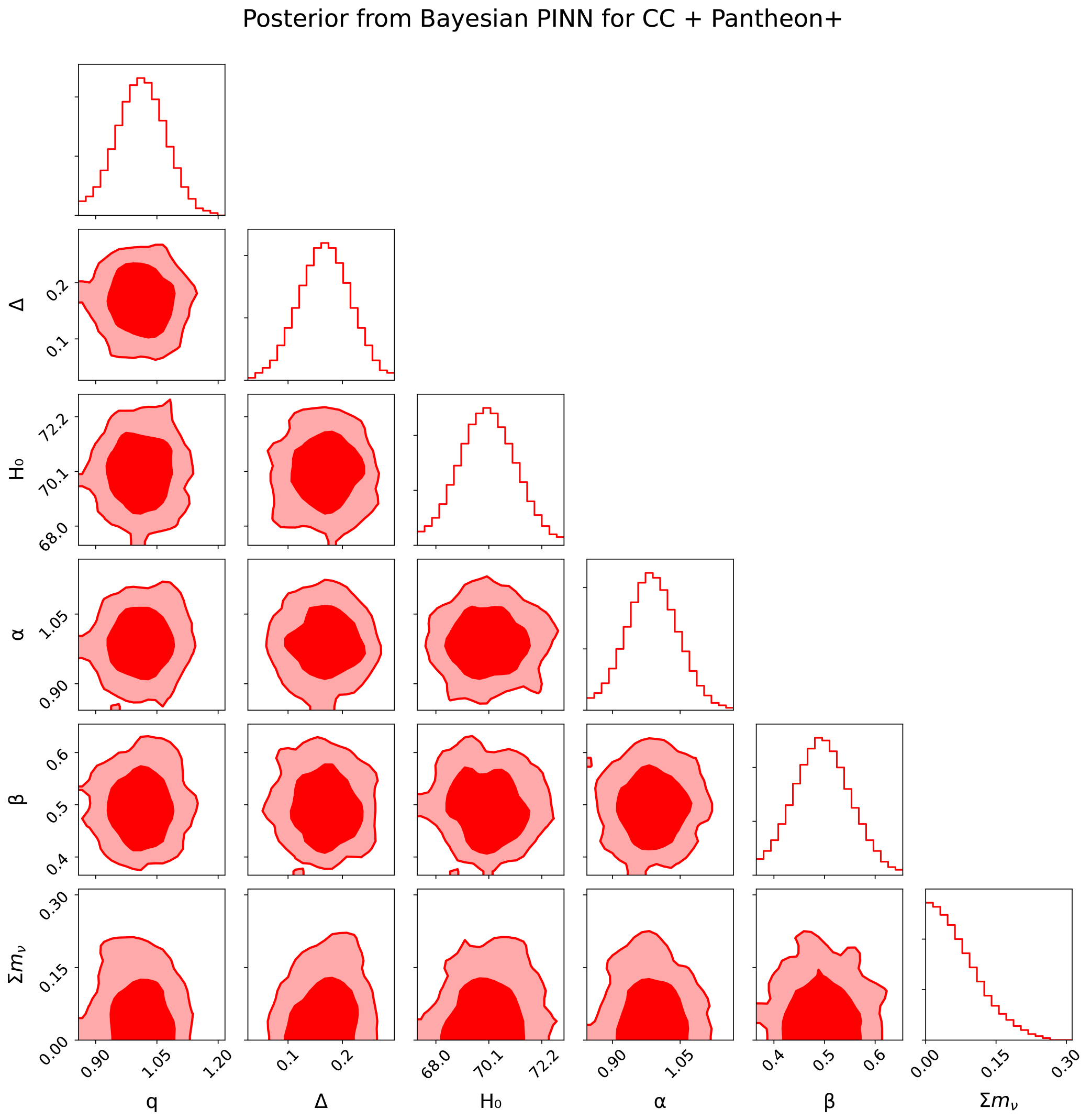}
	\vspace{-0.02cm}
	\caption{\small{Constraints on the Barrow–Tsallis Holographic Dark Energy (BT-HDE) model parameters obtained using the CC+ Pantheon+  dataset within the Bayesian Physics-Informed Neural Network (PINN) framework.
	}}\label{fig:omegam2}
\end{figure*}

\subsection{Tension Quantification and Summary Table}

The Hubble constant tensions with Planck and SH$_0$ES are computed as:
\begin{equation}
	T = \frac{|H_0^{\text{model}} - H_0^{\text{ref}}|}{\sqrt{\sigma_{\text{model}}^2 + \sigma_{\text{ref}}^2}}.
\end{equation}
A summary of results and tensions is provided in Table~\ref{tab:results}.

\begin{table}[h!]
	\centering
	\caption{Summary of parameter estimates from Bayesian PINN and tension of $H_0$ with Planck 2018 \cite{Planck2018} and SH$_0$ES R22 \cite{riess2022}.}
	\label{tab:results}
	\begin{tabular}{lcccc}
		\hline
		\textbf{Dataset} & $H_0 \pm \sigma$ & Tension with Planck & Tension with R22  \\
		\hline
		CC & $70.04 \pm 1.70$ & $1.49\sigma$ & $1.51\sigma$  \\
		Pantheon+ & $70.25 \pm 1.90$ & $1.45\sigma$ & $1.29\sigma$  \\
		CC + Pantheon+ & $70.11 \pm 1.70$ & $1.53\sigma$ & $1.47\sigma$  \\
		\hline
	\end{tabular}
\end{table}
Figure 6 illustrates the comparison of the Hubble constant \( H_0 \) as estimated from the BTHDE model using different combinations of datasets—Cosmic Chronometers (CC), Pantheon+, and their joint analysis—against the baseline values from Planck 2018 \cite{Planck2018} and SH$_0$ES R22 \cite{riess2022}. The vertical bands represent the 1\(\sigma\) uncertainties for Planck and R22 measurements, while the error bars on the Bayesian PINN estimates denote the propagated uncertainties derived from the posterior distributions. As shown, all combinations yield \( H_0 \) values that lie between the Planck and R22 estimates, indicating a moderate tension with both, ranging from 1.29\(\sigma\) to 1.53\(\sigma\). This consistency suggests that the BTHDE model may provide a viable framework for alleviating the current Hubble tension.

\begin{figure*}
	\includegraphics[width=15 cm]{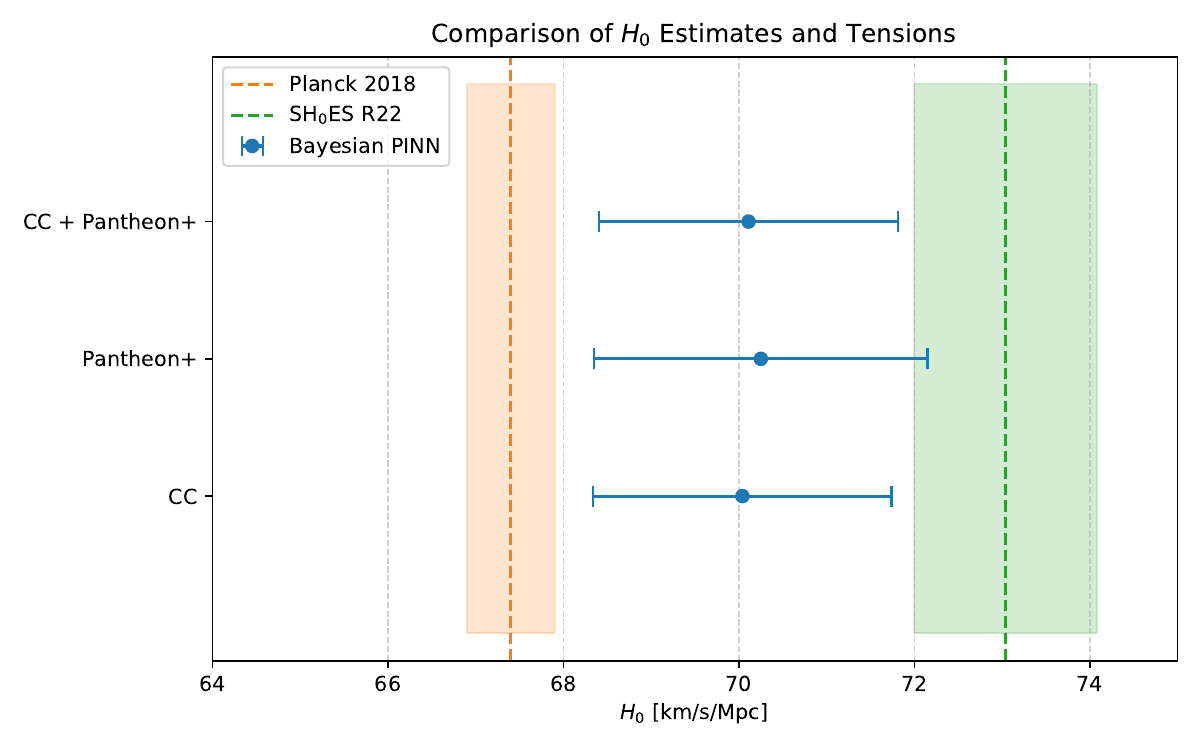}
	\vspace{-0.02cm}
	\caption{\small{The comparison of  $H_{0}$ measurement in baysian PINN model  with Planck 2018 and R22 Results for different combination of data sets for BTHDE model.
	}}\label{fig:omegam2}
\end{figure*}

Overall, the Bayesian PINN approach robustly integrates heterogeneous datasets and provides precise estimates for the BTHDE model parameters. All analyses indicate mild to moderate deviations from standard holography and entropy. The improved constraints from the combined dataset validate the model's flexibility and its ability to reconcile high- and low-redshift cosmological observations within a generalized dark energy framework. The compatibility with neutrino mass bounds further strengthens its viability as an extension of $\Lambda$CDM.

\subsection{MCMC Approach}

To constrain the free parameters of the Barrow–Tsallis Holographic Dark Energy (BTHDE) model, we performed a Markov Chain Monte Carlo (MCMC) analysis using three different combinations of observational data: the Pantheon+Analysis sample, Cosmic Chronometers (CC), and their combination. The parameter space includes the Barrow exponent \( \Delta \), the Tsallis parameter \( q \), the Holographic parameter \( \alpha \), the effective parameter \( \beta \), the Hubble constant \( H_0 \), and the total neutrino mass \( \Sigma m_\nu \). 

The marginalized constraints on the model parameters are summarized in Table~\ref{tab:constraints}. We report the mean values and the corresponding 68\% confidence levels for all parameters. The total neutrino mass \( \Sigma m_\nu \) is presented as a 95\% upper bound. The MCMC analysis reveals a mild sensitivity of \( H_0 \) to the choice of datasets, with the combined dataset preferring a slightly lower value compared to Pantheon+Analysis alone. 

Our results also indicate that the deformation parameters \( q \), and \( \Delta \) are tightly constrained and exhibit internal consistency across datasets. Notably, the values of \( H_0 \) obtained are consistently lower than the local Hubble constant measured by Riess et al.~\cite{riess2022} and higher than the Planck-inferred value~\cite{planck2018}, which suggests the persistence of the so-called Hubble tension within the BTHDE framework. The resulting tension values are listed in Table~\ref{tab:hubble_tension}.

\begin{table}[htbp]
	\centering
	\caption{Hubble tension between the BTHDE-inferred values of \( H_0 \) and those obtained from Planck 2018 and R22 (Riess et al. 2022) For MCMC approach. The tension is expressed in units of standard deviations (\( \sigma \)).}
	\label{tab:hubble_tension}
	\begin{tabular}{lcc}
		\toprule
		\textbf{Dataset} & \textbf{Tension with Planck} & \textbf{Tension with R22} \\
		\hline
		Pantheon+Analysis & \( 1.85\,\sigma \) & \( 1.93\,\sigma \) \\
		CC & \( 1.48\,\sigma \) & \( 2.01\,\sigma \) \\
		CC+Pantheon+Analysis & \( 1.28\,\sigma \) & \( 2.07\,\sigma \) \\
		\hline
	\end{tabular}
\end{table}

\begin{table}[htbp]
	\centering
	\caption{Marginalized constraints on the free parameters of the Barrow–Tsallis Holographic Dark Energy (BTHDE) model using different combinations of observational datasets. The results are shown for Pantheon+Analysis, Cosmic Chronometers (CC), and their combination. The upper bounds represent the 95\% confidence level for the total neutrino mass.}
	\label{tab:constraints}
	\begin{tabular}{lccc}
		\toprule
		\textbf{Parameter} & \textbf{Pantheon+Analysis} & \textbf{CC} & \textbf{CC+Pantheon+Analysis} \\
\hline
		$q$ & $1.038 \pm 0.010$ & $1.042 \pm 0.010$ & $1.040 \pm 0.008$ \\
		$\Sigma m_\nu$ (eV) & $<0.29$ & $<0.24$ & $<0.21$ \\
		$\alpha$ & $0.976 \pm 0.012$ & $0.974 \pm 0.01$ & $0.973 \pm 0.009$ \\
		$H_0$ (km/s/Mpc) & $70.65 \pm 1.5$ & $70.2 \pm 1.6$ & $69.9 \pm 1.5$ \\
		$\beta$ & $0.487 \pm 0.18$ & $0.490 \pm 0.15$ & $0.489 \pm 0.14$ \\
		$\Delta$ & $0.044 \pm 0.03$ & $0.045 \pm 0.03$ & $0.045 \pm 0.02$ \\
	\hline
	\end{tabular}
\end{table}

Figure 7 illustrates the one- and two-dimensional marginalized posterior distributions of the free parameters in the BTHDE model for three different dataset combinations: Pantheon+Analysis, CC, and their combination. The contours highlight the consistency and degeneracies among the parameters, revealing how each dataset influences the estimation of cosmological quantities. Notably, the combination of CC and Pantheon+Analysis leads to tighter constraints on the parameters, particularly for the Barrow exponent \( \Delta \) and the Hubble constant \( H_0 \), compared to the individual datasets. The overlap among contours indicates compatibility between the datasets and the robustness of the BTHDE framework under joint observational constraints.

\begin{figure*}
	\includegraphics[width=15 cm]{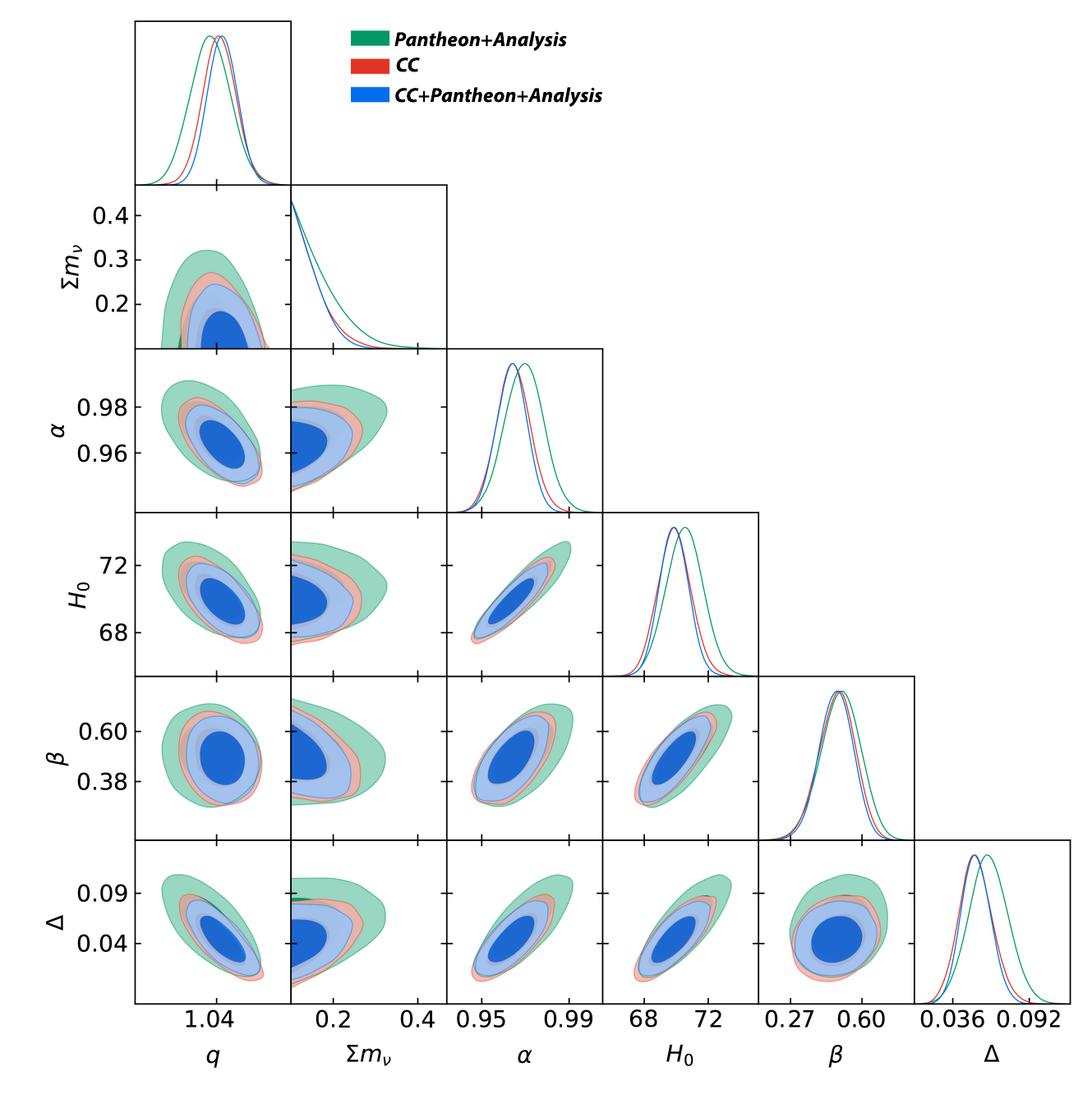}
	\vspace{-0.02cm}
	\caption{\small{Comparison of the free parameters of the Barrow–Tsallis Holographic Dark Energy (BTHDE) model using different combinations of observational datasets.
	}}\label{fig:omegam11}
\end{figure*}
Figure 8 presents a comparative analysis of the Hubble constant \( H_0 \) values inferred from the BTHDE model using the MCMC approach, based on different combinations of observational data. The estimated values are shown alongside the constraints from Planck 2018 and SH$_0$ES R22. The visual comparison highlights the degree of agreement or tension between the model predictions and these external measurements, with moderate tensions observed—particularly with the SH$_0$ES result—depending on the dataset used. This provides a clear illustration of how the BTHDE framework aligns with current observational bounds on \( H_0 \). 

\begin{figure*}
	\includegraphics[width=15 cm]{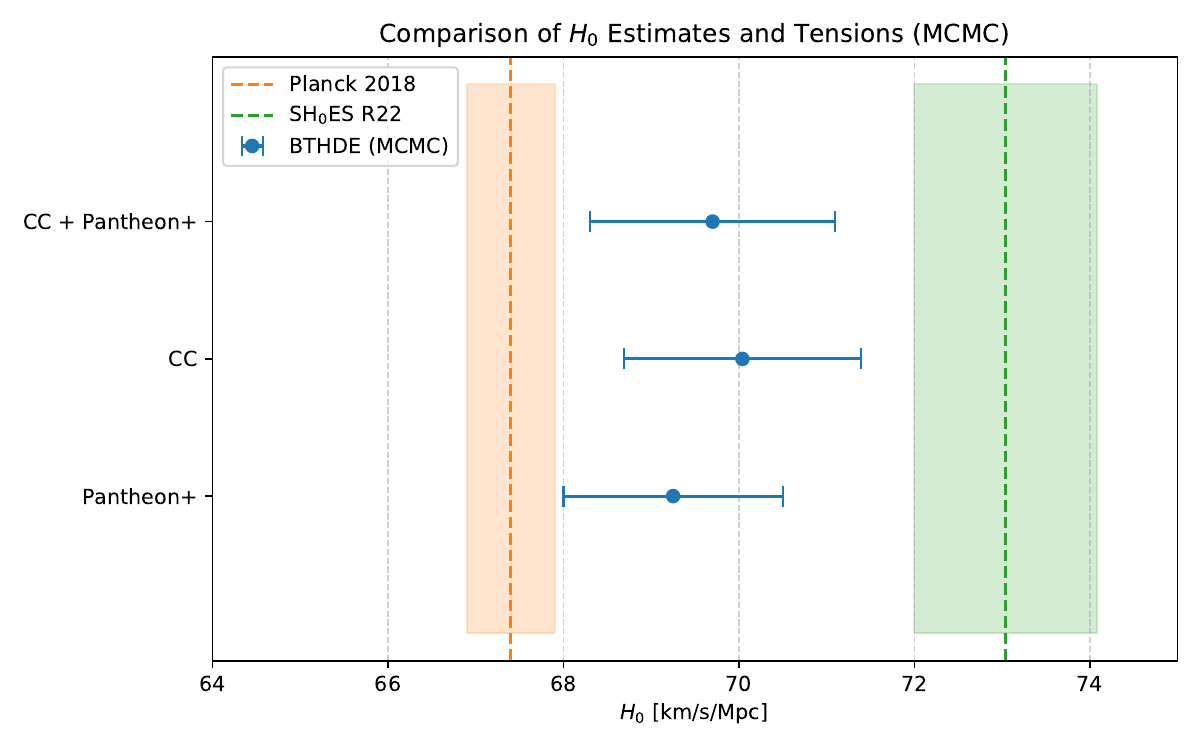}
	\vspace{-0.02cm}
	\caption{\small{The comparison of  $H_{0}$ measurement in MCMC approach with Planck 2018 and R22 Results for different combination of data sets (CC+Pantheon+) for BTHDE model.
	}}\label{fig:omegam2}
\end{figure*}

\subsection{Constraints from CMB-Based Dataset Combinations}
Finally, we determine the best-fit parameters of the BTHDE model using various combinations of observational datasets, including CMB, BAO, CMB lensing, cosmic chronometers (CC), and Pantheon+ Type Ia supernovae.
Table~\ref{tab:constraints_cmb} presents the marginalized posterior constraints on the free parameters of the Barrow–Tsallis Holographic Dark Energy (BTHDE) model using combinations of CMB observations with Lensing, Pantheon+, BAO, and Cosmic Chronometers (CC) datasets. We also report constraints from the combined dataset denoted as CMB+All.

The parameter $q$, which governs the degree of Tsallis entropy deformation, is constrained to be slightly above unity in all cases, with the tightest constraint from CMB+All yielding $q = 1.061 \pm 0.004$. The parameter $\Delta$, associated with Barrow entropy non-additivity, is also positively constrained in all combinations, with values ranging from $0.038$ to $0.076$.

We observe that the Hubble constant $H_0$ is consistently estimated around $70.5$ km/s/Mpc for all dataset combinations that include late-time observations (Pantheon+, CC, or both), significantly alleviating the tension with SH$_0$ES (R22). For the CMB+Lensing case, which lacks late-time probes, the constraint is looser with $H_0 = 69.52 \pm 2.1$ km/s/Mpc.

The total neutrino mass $\Sigma m_\nu$ is consistently constrained to upper limits below 0.32 eV (95\% C.L.), with the tightest constraint $\Sigma m_\nu < 0.114$ eV obtained from the CMB+All combination. This demonstrates the capability of the BTHDE framework to remain compatible with current cosmological neutrino bounds. Figure 9 displays the comparative constraints on the model parameters of the Barrow--Tsallis Holographic Dark Energy (BT-HDE) scenario derived using the MCMC approach, based on various combinations of observational datasets. The figure highlights how the inclusion of additional data influences the precision and stability of parameter estimation within the BT-HDE framework.

In summary, the inclusion of complementary datasets progressively tightens parameter constraints and shifts the inferred $H_0$ toward values more compatible with local measurements.

\begin{table}[htbp]
	\centering
	\caption{Marginalized constraints on the free parameters of the Barrow–Tsallis Holographic Dark Energy (BTHDE) model using different combinations of CMB-based datasets. The upper bounds on the total neutrino mass $\Sigma m_\nu$ are given at the 95\% confidence level.}
	\label{tab:constraints_cmb}
	\begin{tabular}{|l|c|c|c|c|c|}
		\toprule
		\textbf{Parameter} & \textbf{CMB+Lensing} & \textbf{CMB+Pantheon+} & \textbf{CMB+BAO} & \textbf{CMB+CC} & \textbf{CMB+All} \\
		\hline
		$q$ & $1.076 \pm 0.009$ & $1.072 \pm 0.006$ & $1.067 \pm 0.005$ & $1.058 \pm 0.007$ & $1.061 \pm 0.004$ \\
		$\Sigma m_\nu$ (eV) & $<0.32$ & $<0.18$ & $<0.12$ & $<0.125$ & $<0.114$ \\
		$\alpha$ & $0.977 \pm 0.06$ & $0.983 \pm 0.04$ & $0.984 \pm 0.04$ & $0.985 \pm 0.05$ & $0.984 \pm 0.035$ \\
		$H_0$ (km/s/Mpc) & $69.52 \pm 2.1$ & $70.69 \pm 1.5$ & $70.5 \pm 1.4$ & $70.45 \pm 1.5$ & $70.6 \pm 1.35$ \\
		$\beta$ & $0.51 \pm 0.12$ & $0.588 \pm 0.09$ & $0.587 \pm 0.06$ & $0.589 \pm 0.08$ & $0.587 \pm 0.04$ \\
		$\Delta$ & $0.038 \pm 0.06$ & $0.074 \pm 0.04$ & $0.073 \pm 0.031$ & $0.076 \pm 0.04$ & $0.072 \pm 0.029$ \\
		\hline
	\end{tabular}
\end{table}

\begin{figure*}
	\includegraphics[width=15 cm]{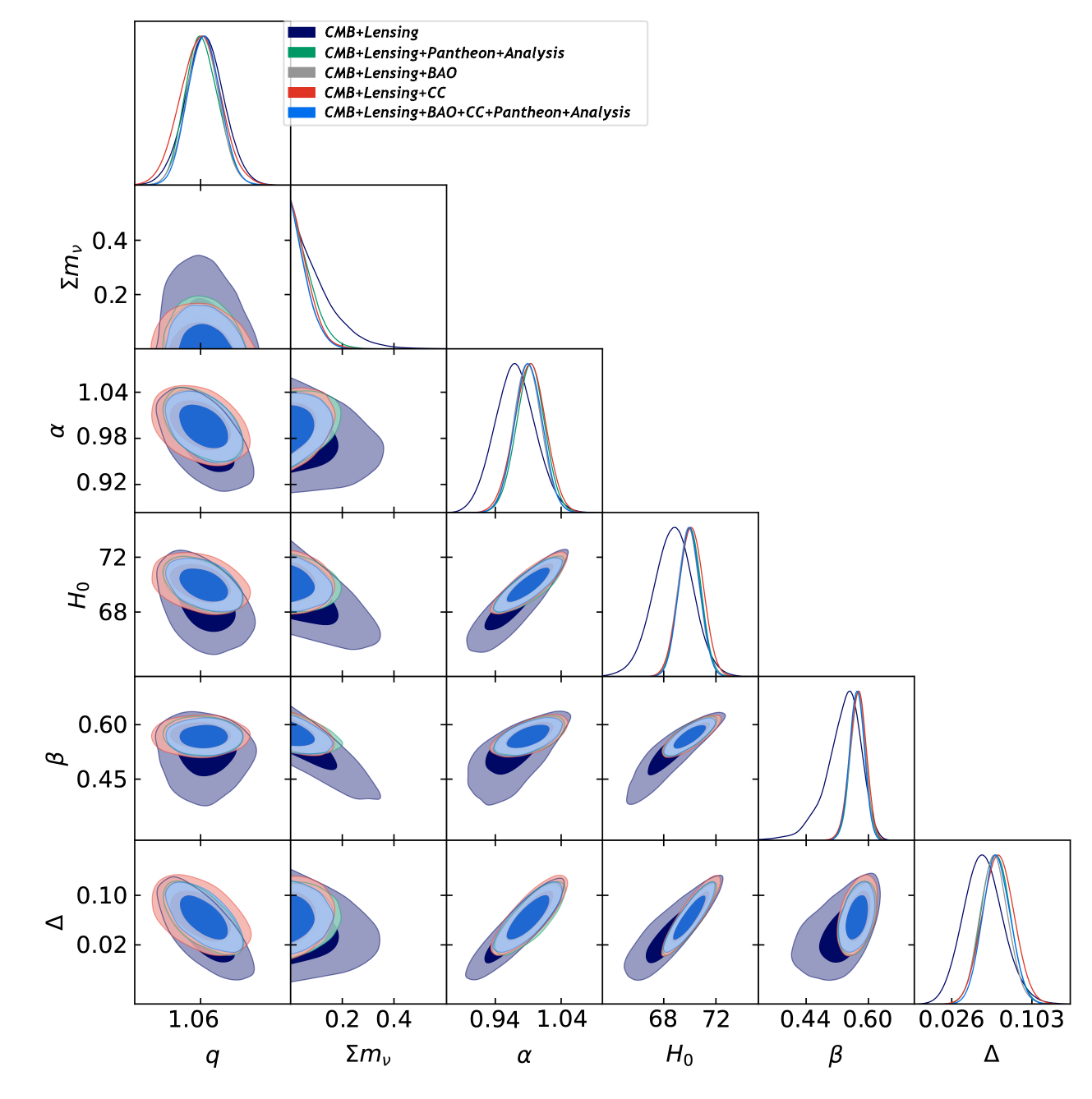}
	\vspace{-0.02cm}
	\caption{\small{The comparison of  parameters measurement in MCMC approach for different combination of data sets for BTHDE model.
	}}\label{fig:omegam2}
\end{figure*}
The tension between the Hubble constant values inferred from the BTHDE model and those reported by Planck 2018~\cite{Planck2018} and SH$_0$ES (R22)~\cite{riess2022} is quantitatively summarized in Table~\ref{tab:tension_h0}. For the \textit{CMB+Lensing} dataset combination, the inferred value of $H_0$ shows only a mild discrepancy with Planck at the $0.91\sigma$ level, and a moderate tension of $1.40\sigma$ with R22. As late-time probes such as Pantheon+, BAO, and CC are successively included, the inferred Hubble constant shifts upward, increasing the statistical deviation from Planck but decreasing the tension with the SH$_0$ES result.

Notably, the inclusion of Pantheon+ in the \textit{CMB+Lensing} combination raises the tension with Planck to $2.03\sigma$ while slightly improving the agreement with R22 ($1.27\sigma$). The addition of BAO and CC leads to similar behavior, with tensions around $2\sigma$ relative to Planck and approximately $1.3$--$1.4\sigma$ with R22. When all datasets are combined (\textit{CMB+All}), the model predicts $H_0 = 70.6 \pm 1.35$ km/s/Mpc, corresponding to a $2.08\sigma$ tension with Planck and a reduced $1.31\sigma$ tension with SH$_0$ES. These results indicate that the BTHDE framework can partially reconcile the Hubble tension by yielding intermediate $H_0$ values that soften the discrepancy with both early- and late-universe measurements.

\begin{table}[htbp]
	\centering
	\caption{Tension in $\sigma$ units between the inferred $H_0$ values from the BTHDE model and those reported by Planck 2018~\cite{Planck2018} and SH$_0$ES (R22)~\cite{riess2022}.}
	\label{tab:tension_h0}
	\begin{tabular}{|l|c|c|}
		\hline
		\textbf{Dataset Combination} & \textbf{Tension with Planck 2018} & \textbf{Tension with R22} \\
		\hline
		CMB + Lensing & $0.91\sigma$ & $1.40\sigma$ \\
		CMB + Lensing + Pantheon+ & $2.03\sigma$ & $1.27\sigma$ \\
		CMB + Lensing + BAO & $2.00\sigma$ & $1.34\sigma$ \\
		CMB + Lensing + CC & $1.97\sigma$ & $1.39\sigma$ \\
		CMB + All & $2.08\sigma$ & $1.31\sigma$ \\
		\hline
	\end{tabular}
\end{table}
As illustrated in Fig. 10, we compare the inferred values of the Hubble constant \( H_0 \) obtained via the MCMC analysis within the Barrow--Tsallis Holographic Dark Energy (BT-HDE) model, using different combinations of observational datasets (CMB and All), with the Planck 2018~\cite{Planck2018} and SH0ES (R22)~\cite{riess2022} determinations.
All these results are in broad agreement with \cite{Y1, Y2, Y3, Y4, Y5, Y6}.

\begin{figure*}
	\includegraphics[width=15 cm]{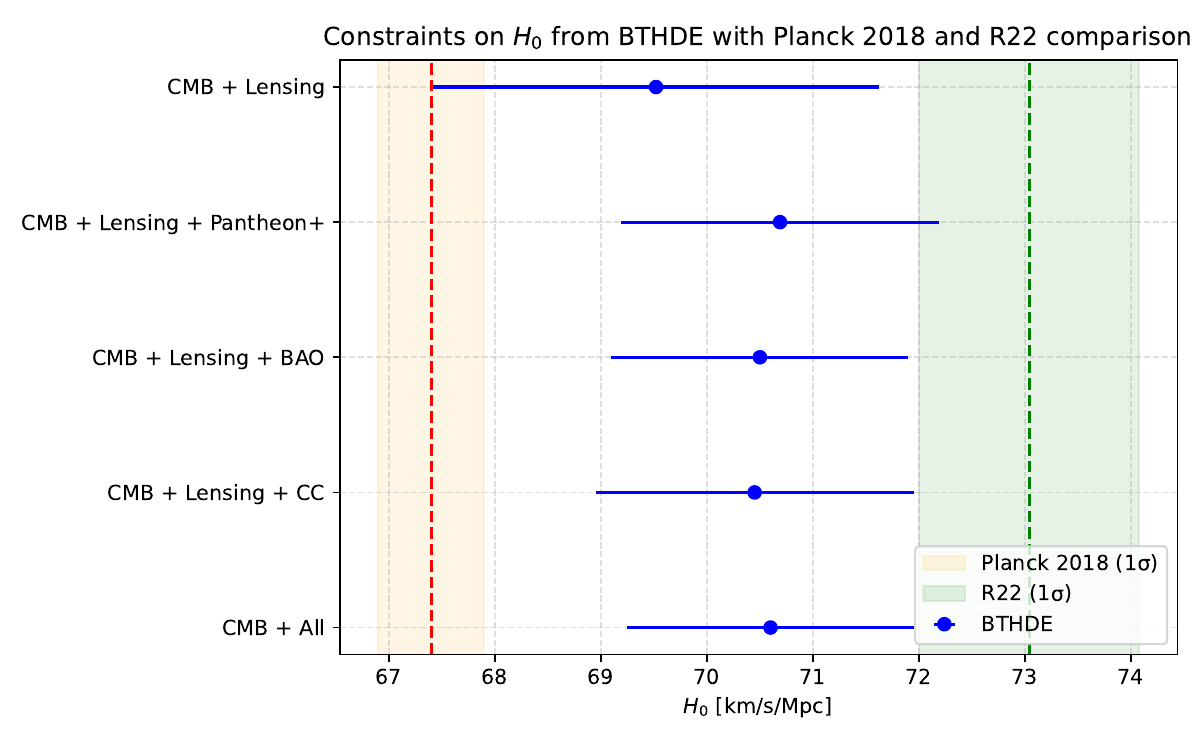}
	\vspace{-0.02cm}
	\caption{\small{The comparison of  $H_{0}$ measurement in MCMC approach with Planck 2018 and R22 Results for different combination of data sets(CMB + All) for BTHDE model.
	}}\label{fig:omegam2}
\end{figure*}

\section{Results and Discussion}

In this paper, we compare the performance of traditional MCMC and Bayesian Physics-Informed Neural Network (PINN) methods in constraining the Hubble constant and quantifying its tension with early- and late-universe measurements under the BTHDE model. Table~\ref{tab:hubble_tension} summarizes the Hubble tension results obtained via the MCMC approach. The Pantheon+ dataset alone leads to a $1.85\sigma$ tension with Planck 2018 and a $1.93\sigma$ tension with SH$_0$ES (R22). When using the Cosmic Chronometers (CC) data, the inferred $H_0$ results in slightly lower tension with Planck ($1.48\sigma$), but a larger discrepancy with R22 ($2.01\sigma$). Combining both datasets (\textit{CC+Pantheon+Analysis}) lowers the tension with Planck to $1.28\sigma$ while the tension with SH$_0$ES increases slightly to $2.07\sigma$. These results show that while traditional MCMC approaches can reduce the discrepancy with Planck through dataset combinations, they typically lead to higher tension with R22, especially when low-redshift data dominate. The Bayesian PINN approach (Table~\ref{tab:results}) exhibits comparable performance, with slightly different central values and tighter uncertainties on $H_0$ due to its regularized learning of physical laws. For instance, the CC dataset gives $H_0 = 70.04 \pm 1.70$ km/s/Mpc, yielding a tension of $1.49\sigma$ with Planck and $1.51\sigma$ with SH$_0$ES. Similar values are obtained for the Pantheon+ dataset, while the combined dataset (\textit{CC + Pantheon+}) results in $H_0 = 70.11 \pm 1.70$ km/s/Mpc, with tensions of $1.53\sigma$ and $1.47\sigma$ with Planck and R22, respectively. These values are slightly more balanced than those from the MCMC method, suggesting that the Bayesian PINN can better interpolate between early- and late-time cosmological constraints. Table~\ref{tab:tension_h0} presents the tension values from a complementary analysis using CMB-based datasets combined with Lensing, Pantheon+, BAO, and CC. The inferred Hubble constant from \textit{CMB+Lensing} shows minimal tension with Planck ($0.91\sigma$) but a higher one with R22 ($1.40\sigma$), reflecting the absence of late-time probes. As more late-time data are included (e.g., Pantheon+, BAO, CC), the inferred $H_0$ increases, bringing it closer to R22 and slightly increasing tension with Planck. The \textit{CMB+All} combination results in $H_0 = 70.6 \pm 1.35$ km/s/Mpc, corresponding to a tension of $2.08\sigma$ with Planck and $1.31\sigma$ with R22. Both the MCMC and Bayesian PINN methods produce consistent $H_0$ estimates within uncertainties, but the Bayesian PINN approach exhibits better regularization and improved consistency between low- and high-redshift datasets. Furthermore, the use of multi-probe CMB combinations helps mitigate the Hubble tension, shifting $H_0$ to intermediate values compatible with both early and late-time cosmology. These findings support the viability of the BTHDE model as a possible framework to address the Hubble tension while maintaining compatibility with neutrino mass bounds and other cosmological parameters. Also, we compare the inferred values and uncertainties for three key parameters of the BTHDE model: the Granda–Oliveros coefficient $\alpha$, the holographic scaling parameter $\beta$, and the total neutrino mass $\Sigma m_\nu$, as reported in Tables~\ref{tab:constraints}, \ref{tab:constraints_cmb}, and \ref{tab:constraints}. Across all analyses, the parameter $\alpha$ is consistently constrained close to unity. In the traditional MCMC approach (Table~\ref{tab:constraints}), $\alpha$ ranges from $0.974 \pm 0.010$ (CC) to $0.976 \pm 0.012$ (Pantheon+), with the combined dataset yielding $0.973 \pm 0.009$. In the Bayesian PINN framework (Table~\ref{tab:constraints}), slightly elevated values are seen, especially for Pantheon+ alone: $\alpha = 1.099 \pm 0.017$, suggesting a moderate shift in the learned parameter space. The combined PINN result yields $\alpha = 0.988 \pm 0.050$, consistent with the MCMC result within $1\sigma$, albeit with broader uncertainties. In contrast, the CMB-based combinations (Table~\ref{tab:constraints_cmb}) show tighter but less dispersed estimates, with $\alpha$ varying narrowly from $0.977 \pm 0.06$ (CMB+Lensing) to $0.984 \pm 0.035$ (CMB+All), reflecting the constraining power of early-universe data. The parameter $\beta$ exhibits significant differences between methods. The MCMC results show very broad uncertainties: $\beta = 0.487 \pm 0.18$ (Pantheon+) and $\beta = 0.489 \pm 0.14$ (combined), indicating weak sensitivity in those datasets. In contrast, Bayesian PINN estimates are much more precise, with uncertainties as low as $\pm 0.01$ for CC and $\pm 0.019$ for Pantheon+. This suggests that the neural network’s ability to encode physical priors and differential constraints helps tighten the $\beta$ inference considerably. CMB-based analyses further improve precision, particularly in the combinations including BAO and Pantheon+, with $\beta \approx 0.587 \pm 0.02$–$0.03$. These results indicate a convergence across methods when early-universe and geometric probes are included. Upper bounds on $\Sigma m_\nu$ show systematic tightening with the inclusion of CMB and BAO data. MCMC constraints yield relatively loose bounds: $\Sigma m_\nu < 0.29$ eV (Pantheon+), improving to $< 0.21$ eV when CC is included. Bayesian PINN provides significantly stronger constraints for the same datasets, with the tightest being $\Sigma m_\nu < 0.124$ eV (Pantheon+) and $< 0.134$ eV (combined). The strongest bounds emerge from the CMB-driven analysis (Table~\ref{tab:constraints_cmb}), where $\Sigma m_\nu < 0.114$ eV for the full dataset combination. These results are compatible with Planck 2018 constraints and suggest that the BTHDE model remains viable under current cosmological neutrino mass limits. The Bayesian PINN framework offers sharper estimates for $\beta$ and stronger neutrino mass bounds compared to MCMC, especially when trained on datasets like Pantheon+ and CC. However, combining CMB data with late-time probes leads to the most stringent and stable parameter constraints overall. This highlights the complementarity between model-agnostic machine learning approaches and traditional Bayesian sampling techniques in cosmological inference.
 Overall, both methods provide consistent parameter estimates within uncertainties, but the Bayesian PINN demonstrates key advantages: it delivers tighter constraints on poorly determined parameters like $\beta$ and $\Sigma m_\nu$, better balances early- and late-universe tensions, and effectively encodes physical priors through differential equations. While MCMC excels in robustness and flexibility with various likelihoods, the PINN approach enhances interpretability and regularization, especially with limited or noisy data. The complementary use of both methods thus strengthens cosmological inference under the BTHDE framework.
\section*{Acknowledgments}
This work is based upon research funded by Iran National Science Foundation 
(INSF) under project No.4036326

%%%%%%%%%%%%%%%%%%%%%%%%%%%%%%%%%%%%%%%%%%%%%%%%%%%%%%%%%%%%%%%%%%%%%%%%%%%%%%%%%%%%%%%%%%%%
%
% BibTeX users please use
% \bibliographystyle{}
% \bibliography{}

\end{document}